\documentclass[iop,apj]{emulateapj}

\slugcomment{ApJ, in press}
\shorttitle{Halo Model Systematics}
\shortauthors{Bailin et~al.}

\defcitealias{bj05}{BJ05}
\defcitealias{cooper-etal10}{C10}
\defcitealias{libeskind-etal11}{L11}

\newcommand{\rvir}{{\ensuremath{r_{\mathrm{vir}}}}}
\newcommand{\meanrho}{{\ensuremath{\left<\rho\right>}}}
\newcommand{\meanm}{{\ensuremath{\left<m\right>}}}
\newcommand{\Msun}{{\ensuremath{M_{\sun}}}}
\newcommand{\df}{{\ensuremath{\mathrm{d}}}}

\begin{document}

\title{Systematic Problems With Using Dark Matter Simulations to Model Stellar Halos}

\author{Jeremy Bailin\altaffilmark{1}, Eric F. Bell\altaffilmark{2},
 Monica Valluri\altaffilmark{2}, Greg S. Stinson\altaffilmark{3},
 Victor P. Debattista\altaffilmark{4}, H. M. P. Couchman\altaffilmark{5}
 and James Wadsley\altaffilmark{5}}
\altaffiltext{1}{Department of Physics and Astronomy, University of Alabama,
  Box 870324, Tuscaloosa, AL, 35487-0324; jbailin@ua.edu}
\altaffiltext{2}{Department of Astronomy, University of Michigan, 830 Dennison Building,
  500 Church Street, Ann Arbor, MI 48109}
\altaffiltext{3}{Max-Planck-Institut f\"ur Astronomie (MPIA), K\"onigstuhl 17, D-69117 Heidelberg, Germany}
\altaffiltext{4}{Jeremiah Horrocks Institute, University of Central Lancashire, Preston, PR1 2HE, UK}
\altaffiltext{5}{Department of Physics and Astronomy, McMaster University, 1280 Main Street West,
  Hamilton, ON L8S 4M1, Canada}

\begin{abstract}
The limits of available computing power have forced models for
the structure of stellar halos to adopt one or both of the following
simplifying assumptions: (1) stellar mass can be ``painted'' onto
dark matter particles in progenitor satellites;
(2) pure dark matter simulations that do not form a luminous
galaxy can be used. We estimate the magnitude of
the systematic errors introduced by these assumptions using a controlled
set of stellar halo models where we independently vary whether we look
at star particles or painted dark matter particles, and whether we use a
simulation in which a baryonic disk galaxy forms or a matching pure dark
matter simulation that does not form a baryonic disk.
We find that the ``painting'' simplification
reduces the halo concentration and internal structure,
predominantly because painted dark matter
particles have different kinematics than star particles even when both
are buried deep in the potential well of the satellite. The simplification
of using pure dark matter simulations
reduces the concentration further,
but \emph{increases} the internal structure,
and results in a more prolate stellar halo.
These differences
can be a factor of $1.5$--$7$ in concentration (as measured by the half-mass
radius) and $2$--$7$ in internal density structure. Given this level of
systematic uncertainty, one should be wary of overinterpreting
differences between observations and the current generation of stellar halo
models based on dark matter only simulations
 when such differences are less than an order of magnitude.
\end{abstract}


\section{Introduction}

While it is now abundantly clear that much of the mass in the extended outer stellar envelopes
of galaxies (stellar halos hereafter) is stripped from dwarf galaxies as they tidally interact
with the central galaxy (\citealp{majewski-etal03}; \citealp{bj05}, hereafter \citetalias{bj05};
\citealt{purcell-etal07}; \citealp{bell-etal08,mcconnachie-etal09};
\citealp{cooper-etal10}, hereafter \citetalias{cooper-etal10}; \citealp{xue11,ibata13}),
a number of questions remain.
Is \textit{all} halo mass accreted (\citetalias{bj05,cooper-etal10}; \citealp{rashkov12}),
or are substantial fractions
kicked up from the stellar disk (called \textit{in situ}; \citealp{kazantzidis08,zolotov09,font-etal11})
or formed within satellites after they have been accreted (\citealp{tissera13}; Valluri et al., in prep.)?
How much variation is expected from halo to halo (\citetalias{bj05}, \citealp{bell-etal08,font-etal11})?

Many studies focus on
these issues by comparing observations of stellar halos with models of stellar halo formation
in a cosmological context \citep{bell-etal08,bell-etal10,helmi-etal11,xue11,schlaufman12,monachesi13}.
In most cases, cognizant that much of the stellar halo is accreted, comparisons are made to models
in which the stellar halo is constituted \textit{only} of accreted material: this simplifies
interpretation, and any major discrepancies between the observations and such models could
signal that this assumption is incorrect, giving insight into other possible modes of halo formation.

A generic practical problem that one encounters when creating a theoretical numerical
model of stellar halos is resolution. Although stellar halos are 
potentially rich with
signposts of the hierarchical galaxy formation process, they contain a very
small fraction of the total stellar mass in a galaxy. For example, the stellar halos of the
Milky Way and similar-mass galaxies
account for only $\approx 1$ -- $10\%$ of their stellar content
(\citetalias{bj05}; \citealp{bailin11,ibata13}).
In order to
resolve the tidal streams that constitute the halo and provide observational tests
of the hierarchical merging paradigm, hundreds of thousands of particles
must be used within the halo itself. Simulating the entire galaxy self-consistently at
this resolution would then
require hundreds of millions of particles, a task that would require tens of millions
of CPU hours per galaxy with current algorithms and hardware. The problem
is exacerbated by the stochastic nature of merger histories, which result in factors of
several galaxy-to-galaxy variation in stellar halo properties even at a given galaxy
mass \citep[e.g.][]{purcell-etal07}
and requires performing a significant number of these simulations
in order to make robust predictions.

The common solution to this problem is to not simulate the entire galaxy self-consistently.
Pure $N$-body simulations are much faster than full hydrodynamic simulations at a given
resolution, at the expense of not including any non-gravitational processes
(e.g. \citetalias{cooper-etal10}; \citealp{libeskind-etal11}, hereafter \citetalias{libeskind-etal11}).
Furthermore,
if the main body of the galaxy is replaced by an analytic potential, the number of particles
gets dramatically reduced and it becomes feasible to simulate the halo at high resolution
in a reasonable length of time \citepalias[e.g.][]{bj05}.
In these pure $N$-body methods (\citetalias{cooper-etal10,libeskind-etal11}),
dark matter (DM) particles must be labelled (``painted'') to
represent the stellar component in lieu of having a self-consistent method of generating
stars.
Other tactics that have been taken are to
use lower resolution self-consistent simulations where the internal structure of the halo
is poorly resolved \cite[e.g.][]{font-etal11} (in low resolution simulations,
unresolved physical processes may also have a significant impact on the derived
halo properties, although in the case of \citealp{font-etal11} the authors have tested that their conclusions
are robust to a factor of $2$ change in spatial resolution),
or to use semi-analytic prescriptions that have explicitly
no internal structure but only predict the total amount of halo material
\citep{purcell-etal07}.

Yet, there are important differences between the expectations of these different approaches.
The difference that inspired this work was the factor-of-two difference in the degree of
substructure predicted by \citetalias{bj05} and \citetalias{cooper-etal10}. As described by
\citet{helmi-etal11} and \citet{bell-etal08}, these models predict
an amount of substructure (measured using the RMS of the model around a smooth halo
profile, divided by the total number of stars) different by a factor of two or more
from each other, in the sense that the \citetalias{cooper-etal10} models have considerably
more substructure than those of \citetalias{bj05}
(\citealt{schlaufman12} find a similar difference between the \citetalias{bj05} and
\citealt{rashkov12} models).
The \citetalias{cooper-etal10} models also have more substructure than the observations;
\citet{helmi-etal11} interpreted this as a model--data discrepancy signaling the need for
halo stars to form \textit{in situ}; we interpret this as a \textit{model--model}
discrepancy signaling the need to better understand why two seemingly reasonable models should
disagree so significantly. One potential point of distinction between the models was that, while
\citetalias{bj05} had an analytic potential (and therefore would have been perhaps more
likely to have a more structured halo), it included the potential from a \textit{disk}.

As discussed earlier, the inclusion of baryons in simulations of galaxy formation adds a degree
of complexity and computational cost that both reduces the general applicability of the simulations
and prohibits the construction of samples of stellar halos that adequately span the range of
possible assembly histories. In this work, we use a hydrodynamical simulation from the
McMaster Unbiased Galaxy Simulations project \citep{MUGS}, which is of sufficiently high resolution
that the satellites whose accretion we wish to follow are well resolved,
to explore two crucial aspects of the relationship of baryons to dark matter relevant to stellar halo
formation. Firstly, the dissipative formation of a disk changes the potential of the galaxy,
enhancing the strength of the tidal field and affecting the orbits of halo stars \citep{penarrubia10_cusp}.
Secondly, we wish to explore the importance of the practice of ``painting'' stars onto dark matter particles:
inasmuch as dark matter particles have not suffered dissipation, even the most bound dark matter particles
have orbits that are very likely to be different from realistic stellar orbits, and this would
affect the properties of the resulting stellar halos.

The plan of this paper is as follows.
In Section~2, we give an overview of the assumptions that have been used in previous work and the
models that we will use to test their effects. In Section~3, we provide full details of
how the simulations and models are generated. Section~4 contains the results from
the different models, and in Section~5 we discuss what these results imply about the
influence of the standard assumptions on halo models. Finally, our conclusions are
presented in Section~6.

\section{Overview}
There are two key assumptions that previous halo models have often adopted in order
to make the problem tractable, which we label ``painting'' and ``dark matter dynamics.''

\begin{description}
\item[\textit{Painting}]\label{overview-painting}
All high resolution stellar halo models consist of pure $N$-body simulations that
contain only DM particles.
In order to predict the properties of the luminous
stellar halo, the authors ``paint'' stellar mass onto certain DM particles,
and then measure
the properties and structure of these painted particles. The methods used to paint vary:
\citetalias{bj05} resolve each contributing subhalo into $10^5$ DM particles, and then
paint the most bound particles such that the luminosity follows a King profile;
\citetalias{libeskind-etal11} has equal-mass DM particles that are
painted equally if they are sufficiently deep in the potential well;
\citetalias{cooper-etal10} use a sophisticated
semi-analytic galaxy formation model to determine the expected amount and distribution of
star formation within each contributing subhalo and paint stellar masses onto the DM
particles so as to contain the appropriate star formation history;
and \citet{rashkov12} paint the most-bound DM particles within each subhalo
equally, but with a stellar mass that varies between subhalos.

\item[\textit{Dark Matter Dynamics}]
When galaxies form, baryons cool and collapse into a centrifugally-rotating disk, whose
morphology can then be altered by further accreted material and interactions with other
galaxies; these processes do not occur in pure DM simulations, which do not have the capability
to radiatively cool. The gravitational potential in which the stars that constitute the
stellar halo orbit is therefore different in the real universe than in a pure DM simulation:
it is more concentrated, and is flattened in the inner regions due to the disk
\citep{kazantzidis-etal04,bailin-etal05-diskhalo,debattista08,tissera10}.
Different groups have taken different approaches to account for this effect:
\citetalias{cooper-etal10} and \citet{rashkov12} use pure $N$-body cosmological simulations and neglect
any changes in the potential due to baryonic physics;
\citetalias{bj05} grow an analytic disk potential inside an analytic growing halo
potential;
and finally, a particularly interesting approach is that of
\citetalias{libeskind-etal11}, who compare a full hydrodynamic
cosmological simulation to the identical DM-only simulation using the same initial conditions.
They find that, when a gravitational-potential-based painting scheme is adopted (see above),
the radial distribution of the stellar halo in both simulations is identical, and therefore
argue that the effect of the baryonic physics can be taken entirely into account by
the appropriate painting scheme.
\end{description}

These two types of assumptions have remained largely untested, and their effect on the final
properties of the predicted stellar halos are therefore unknown. Our goal is to use a
set of control simulations in which we vary either the method by which we determine where
``stars'' lie in the simulation volume or the potential in which the particles orbit.
To do this, we compare a full Smoothed Particle Hydrodynamics (SPH) simulation of galaxy formation
from the McMaster Unbiased Galaxy Simulations\citep[MUGS;][]{MUGS} with a DM-only
simulation of the same initial conditions.
We analyze 4 different models for the stellar halo that is formed:
\begin{enumerate}
 \item SPH-STARS: the stars that form self-consistently in the SPH simulation that
 are accreted from satellites;
 \item SPH-PAINTED: DM particles in the SPH simulation that are painted to match
 the mean stellar mass-DM mass relation of satellites in the simulation;
 \item SPH-EXACT: DM particles in the SPH simulation that are painted to match
 the stellar mass of each individual satellite that contributes to the halo; and
 \item DM-PAINTED: painted DM particles in the DM-only simulation.
\end{enumerate}
Full details of how each of these models is constructed is given
in Section~\ref{sec:modelhalos}.
Our painting schemes are calibrated using the luminous satellites within the
SPH simulation. This allows us to directly compare the painted stars with those
that form self-consistently in the simulations, since the same objects should
have the same stellar content.
The comparison between SPH-STARS and SPH-PAINTED halos isolates the effect of
using painted DM particles instead of stars
(and the SPH-EXACT halo can be used to determine what aspect of the painting
scheme is responsible for any differences),
while the comparison
between the SPH-PAINTED and DM-PAINTED halos isolates the effect of the baryonic contribution
to the gravitational potential.

We emphasize here that although we will use the SPH-STARS halo as the reference model,
this is not because we think it is a good approximation to stellar halos in the
real universe. Galaxy formation simulations at the resolution of MUGS generically form
too many stars by a factor of $\sim 2$ \citep{MUGS},
contain too large a fraction of their stellar mass in their spheroid
(for example, MUGS galaxies have a mean bulge fraction of $73\%$ compared to
an observed value of $\sim 40\%$ for comparable-luminosity observed galaxies;
\citealp{tasca11}),
and cannot resolve the majority of streams that constitute the halo.
However, the relative
comparison between the models is valid: if we start with the same amount of stellar
material in the same satellites, then the stellar halos should be similar if the
assumptions we are testing are appropriate.

\section{Simulations}

\subsection{MUGS}
The simulations we analyze are (1) g15784 from the McMaster
Unbiased Galaxy Simulations \citep{MUGS},
and (2) a simulation with the same initial conditions but evolved purely using collisionless
$N$-body dynamics, i.e. only with DM.
These two simulations provide everything we need to cleanly measure
the importance of the assumptions we are testing.
The total mass of the galaxy within the virial radius (i.e. including subhalos)
is $1.4 \times 10^{12}~\Msun$ at $z=0$, of which $1.1 \times 10^{11}~\Msun$
is in the form of stars and $1.0 \times 10^{11}~\Msun$ is in the form of gas.
The simulation uses a $\Lambda$CDM cosmology with $H_0=73~\mathrm{km~s^{-1}~Mpc^{-1}}$,
$\Omega_m=0.24$, $\Omega_{\Lambda}=0.76$, $\Omega_b=0.04$, and $\sigma_8=0.76$
\citep{WMAP3}.
The DM particle mass is
$1.1 \times 10^{6}~\Msun$ in the SPH simulation and $1.3\times 10^6~\Msun$ in the DM-only simulation,%
\footnote{The difference is because both simulations contain the same number of DM particles,
but in the SPH simulation they account for only $83\%$ of the total mass instead of the
entire mass in the DM simulation.}
the initial gas particle mass is $2.2 \times 10^5~\Msun$, and the initial star particle
mass is $6.3 \times 10^4~\Msun$. The gravitational softening is $312.5$~pc.
The visible galaxy that forms at $z=0$
has a prominent disk that can be traced to $10$~kpc in gas and $20$~kpc in stars,
with a scale length of $1.7$~kpc and has a bulge-to-total ratio of $0.48$ within $25$~kpc.
It is therefore a good analog for an early-type disk galaxy.

87 snapshots of the SPH simulation were saved, while 62 snapshots of the DM simulation
were saved. DM snapshots were spaced equally in time with
$\approx215$~Myr between snapshots; SPH snapshots exist at all of these times plus some
additional times corresponding to convenient redshifts.

\subsection{Merger Trees}

Halos were found in the simulations using Amiga's Halo Finder \citep[AHF;][]{AHF},
which generates a spatially-adaptive mesh on which the density field is measured,
and then structures are found in the density field corresponding to a virialized
spherical overdensity criterion. Because of the adaptive mesh, structures can be
found on different scales, and so AHF generates a hierarchy of halos, subhalos,
subsubhalos, etc. Finally, energetically unbound particles are removed from the
particle lists corresponding to each halo. The output contains the list of particles
associated with each halo in each snapshot.

To follow the evolution of individual halos, we must associate
dark matter halos in successive snapshots. To do this, we compare the list
of particles IDs
between each pair of halos, and assign halo $h_i$ in snapshot $s_i$ to be the
progenitor of halo $h_{i+1}$ in snapshot $s_{i+1}$
if $h_i$ contributes more particles to $h_{i+1}$ than any other halo in $s_i$ does.
We also must be careful with substructure:
AHF can assign particles to multiple halos, and in particular, the particles of a subhalo
are usually also members of the parent halo. Therefore, if two subhalos merge within
a parent halo, the parent halo may appear to be the progenitor of the merger product,
because it contributes all of the particles in both subhalos.
We therefore assign the progenitor to be the subhalo whose contribution to $h_{i+1}$
most closely matches the actual number of particles in $h_{i+1}$ in these cases.

Linking together each halo with its successor results in a track, which we
consider to be the evolution of an individual object. The merger tree
``trunk'' is the track that results in the parent halo at $z=0$. A list of
the maximum mass each track achieves and the snapshot at which that mass
is reached is recorded and used for the particle painting (Section~\ref{sec:paintingmethod}).

\subsection{Model Halos}\label{sec:modelhalos}
\subsubsection{Accreted Stars}
The SPH-STARS halo model consists of all accreted stars in the full SPH simulation. We
define accreted stars as those that are not within the AHF particle list
(i.e., outside of the virial radius) of the parent
halo trunk at the first simulation output in which they appear, but which appear within
the particle list of the parent halo at $z=0$.
Note that this neglects stars that form in satellites after they have been accreted
into the parent halo, but which are later stripped (\citealp{tissera13}; Valluri et al., in prep.).
Although it is theoretically possible
for a star particle to form in a satellite outside of the trunk but fall into the parent halo
before the next output, outputs are placed closely enough together in time
(at most $215$~Myr) that the number of such particles should be small.
Each star particle is born with mass $6.3 \times 10^4~\Msun$, and then loses mass over time due to stellar
evolution.

The final stellar halo in the SPH-STARS model contains $3.49 \times 10^{10}~\Msun$ in
$799248$ particles within $1.1$ times the virial radius.
Because some observational analyses specifically avoid regions with
known satellites, we also construct a model halo where we exclude all particles contained
in bound substructures that are found by AHF. This ``NOSUBS'' halo model contains
$1.78 \times 10^{10}~\Msun$ in $409645$ particles.

\subsubsection{Painting Methods}\label{sec:paintingmethod}
\begin{figure}
\plotone{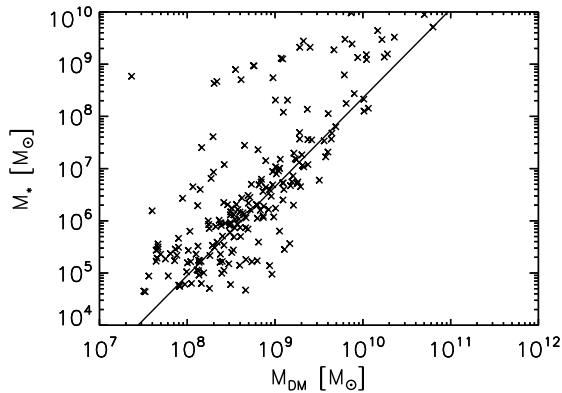}
\caption{\label{fig:mstarcalib}%
The stellar mass $M_*$ of all halos in the MUGS SPH simulation as a function of
their dark matter content $M_{\mathrm{DM}}$ at their time of maximum mass. The
line is the power law fit given in equation~(\ref{eq:mpaint}).}
\end{figure}

All halo models aside from SPH-STARS consist of dark matter particles (in either
the DM-only or SPH simulation) that have
been assigned a stellar mass. This stellar mass is assigned to merger tree
tracks at their maximum-mass snapshot.

\paragraph{SPH-PAINTED}
The SPH-PAINTED model is constructed in the full SPH simulation from DM particles that have
been painted.

In the SPH simulation, we can measure the actual stellar mass fraction for
DM halos at the snapshot where they reach their maximum mass (Figure~\ref{fig:mstarcalib}).
The following relationship provides a good fit to the majority of halos:
\begin{equation}\label{eq:mpaint}
M_* = 4.5 \times 10^6~\Msun  \left(\frac{M_{\mathrm{DM}}}{10^9~\Msun}\right)^{1.7}.
\end{equation}

It is possible in the real universe
to use the observed satellite galaxy luminosity function and the
simulated dark matter subhalo mass function to estimate the stellar mass fraction
satellites must have at infall (which usually corresponds to their time of
maximum mass). Studies that do this require there to be a much steeper
relationship with lower normalization than what is seen in equation~(\ref{eq:mpaint}),
with power law indices ranging from $2.5$ to $3$ and normalizations at $10^9~\Msun$
ranging from $2 \times 10^3$ to $2 \times 10^4~\Msun$ \citep{koposov09,kravtsov10,rashkov12}.
The SPH simulations, however, produce many more stars in the small objects as a consequence
of their well documented tendency to overcool \citep[e.g.][]{MUGS}.
In order to compare the SPH-STARS and SPH-PAINTED model halos, and therefore
determine the effect that particle tagging has on the predicted properties of halos,
we must make the progenitors in these models as similar as possible.
We therefore match the stellar mass of the progenitor to what forms in the hydrodnamic
simulation rather than the stellar content of real galaxies.
If the SPH simulations could perfectly reproduce satellite galaxies, then these
calibrations would be identical.
Equation~(\ref{eq:mpaint}) is therefore the correct stellar mass fraction to adopt
for this purpose.

There are a minority of AHF groups ($23\%$) with an unusually high stellar mass content,
with stellar masses greater than or equal
to their dark masses. These objects are usually subgroups of larger galaxies, and turn
out to be stellar concentrations (e.g., star clusters and spiral arms)
that are not independent contributors to the merger
history of the main galaxy and should not be painted separately.
However, they are assigned a stellar mass that is commensurate with their relatively small dark
mass, and are therefore also small and do not contribute significantly to the 
model halo.

We paint the most bound\footnote{In determining the ``most bound'' particles, we use the
total potential plus kinetic energies.}
$1\%$ of the particles in the halos at the time of maximum mass,
and divide the stellar mass evenly amongst these particles.
We tested painting different fractions of most bound particles,
and found little qualitative difference in the
properties of the resulting stellar halos for values less than $\sim 10\%$.
Our approach is similar to what \citet{rashkov12} adopted to generate
the halo analyzed by \citet{schlaufman12}.

The final SPH-PAINTED halo model contains $2.25 \times 10^{10}~\Msun$ in $5179$ particles,
while the NOSUBS version contains $1.58 \times 10^{10}~\Msun$ in $4043$ particles.

\paragraph{DM-PAINTED}
The DM-PAINTED model is constructed in the DM-only simulation from painted particles.
In order to compare the DM-PAINTED and SPH-PAINTED model halos, and therefore determine the
effect that non-gravitational physics has on the predicted properties of stellar halos,
we must use the identical painting scheme. We therefore also paint the most bound $1\%$
of the particles in the DM halos at the time of maximum mass with a total
stellar mass from equation~(\ref{eq:mpaint}), evenly divided amongst the particles.
The AHF groups that are bound by their stellar mass, rather than their DM, are absent in the DM-only
simulation, and therefore do not contribute at all to the model halo.

The DM-PAINTED halo model contains $1.45 \times 10^{10}~\Msun$ in $4405$ particles,
while the NOSUBS version contains $1.22 \times 10^{10}~\Msun$ in $3862$ particles.

\paragraph{SPH-EXACT}
The SPH-EXACT model, like the SPH-PAINTED model, is constructed in the SPH simulation from
painted DM particles. However, rather than using equation~(\ref{eq:mpaint}) to determine
the total stellar mass of each progenitor, the exact stellar mass for that progenitor
in the SPH simulation is used, as shown in Figure~\ref{fig:mstarcalib}. This allows us
to ascertain whether any differences between SPH-STARS and SPH-PAINTED are
due to the particular method of assigning stellar masses to DM halos,
or whether they are generic to the enterprise of painting DM particles.

The SPH-EXACT halo model contains $6.42\times 10^{10}~\Msun$ in $5179$ particles,
while the NOSUBS version contains $4.49 \times 10^{10}~\Msun$ in $4043$ particles.
The choice of painted particles are identical to the SPH-PAINTED model, but
they are assigned different stellar masses.
The total mass is higher than in SPH-PAINTED because the subhalos that do not fall on the
best fit relation of Figure~\ref{fig:mstarcalib} scatter systematically high.

\section{Halo Structure}

\subsection{Overview}


\begin{figure*}
$\begin{array}{cc}
 \begin{array}{cc}
 \includegraphics[scale=0.65]{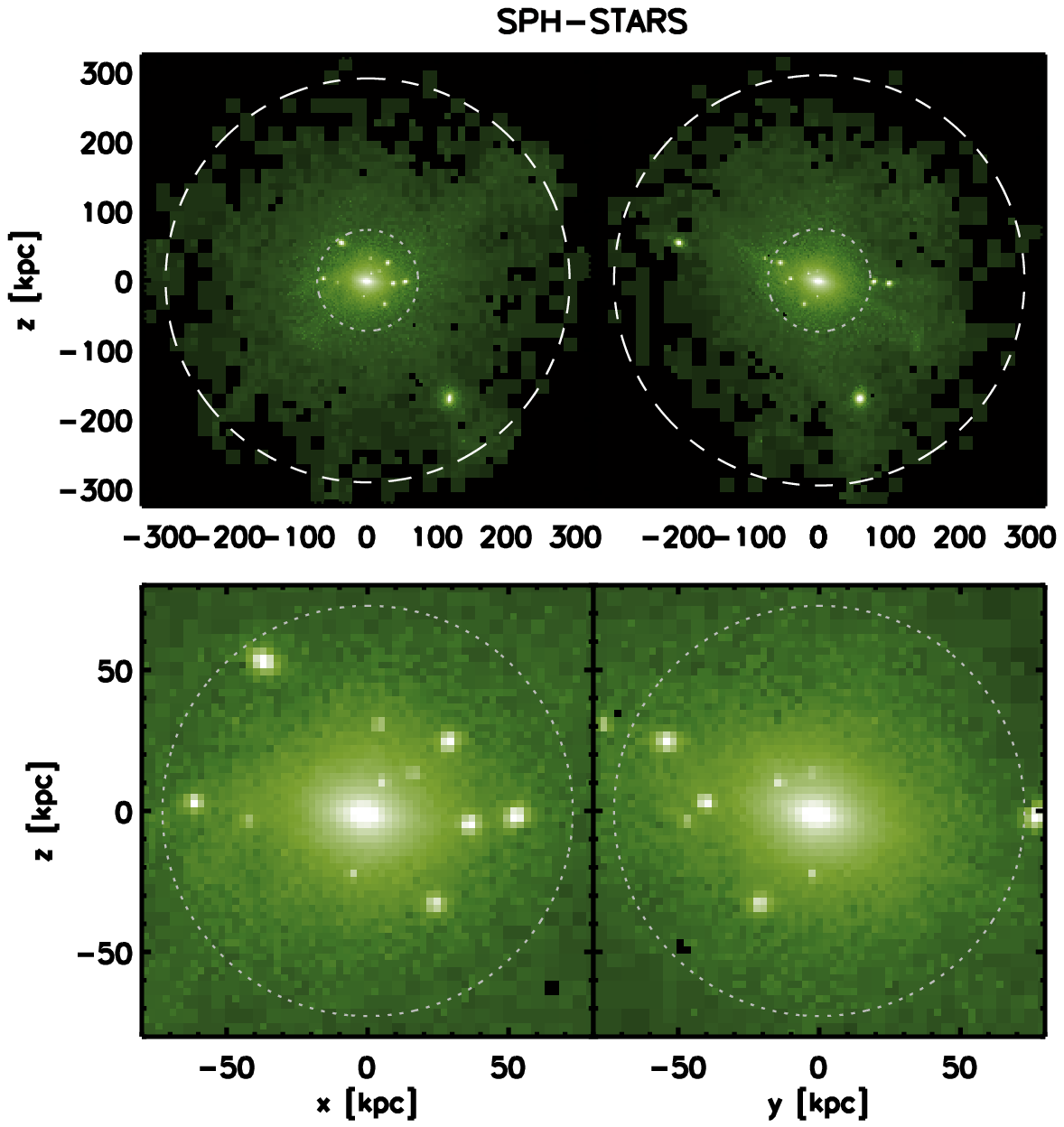} & 
   \includegraphics[scale=0.65]{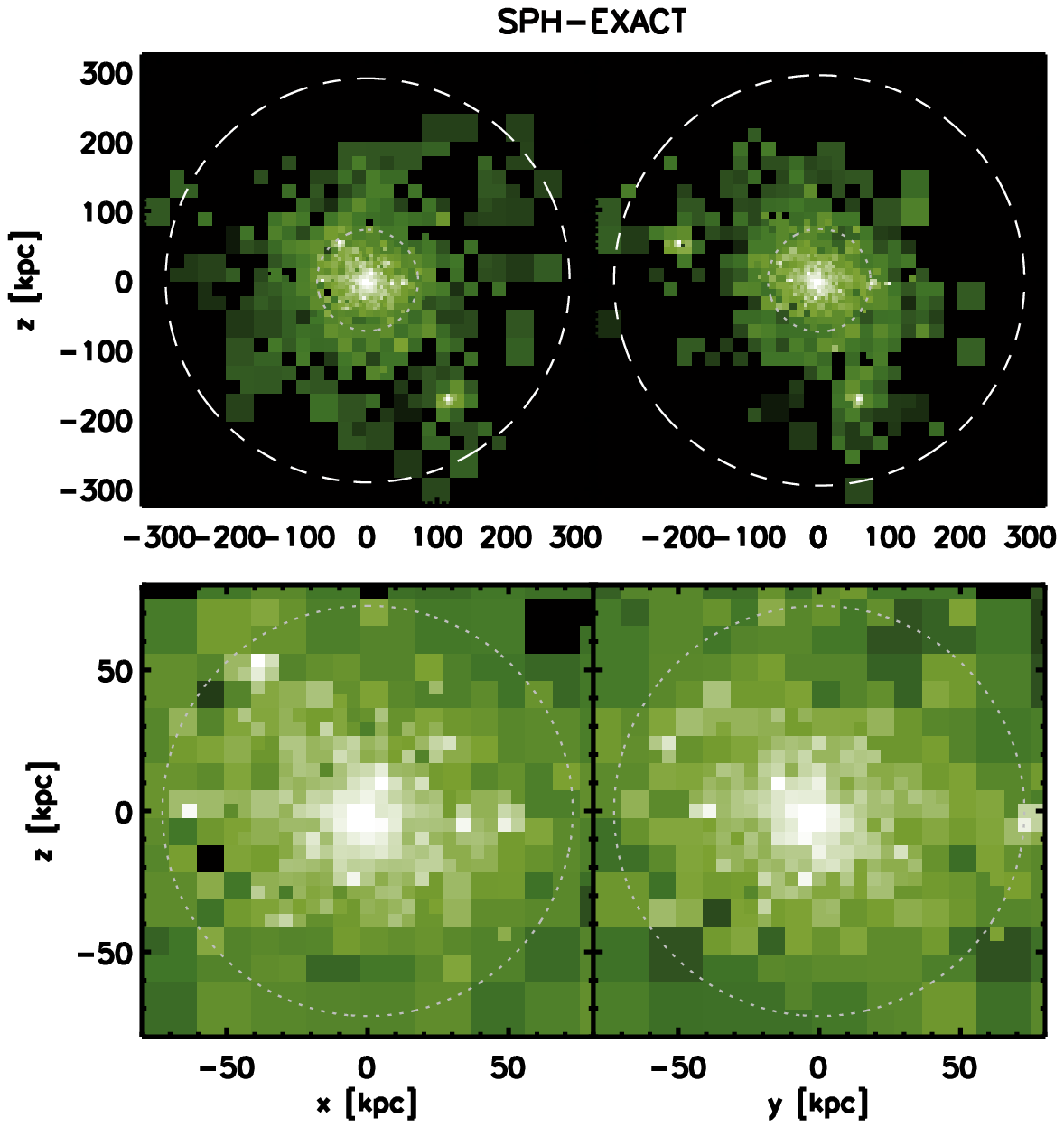} \\
 \vspace{0.5em}\\
 \includegraphics[scale=0.65]{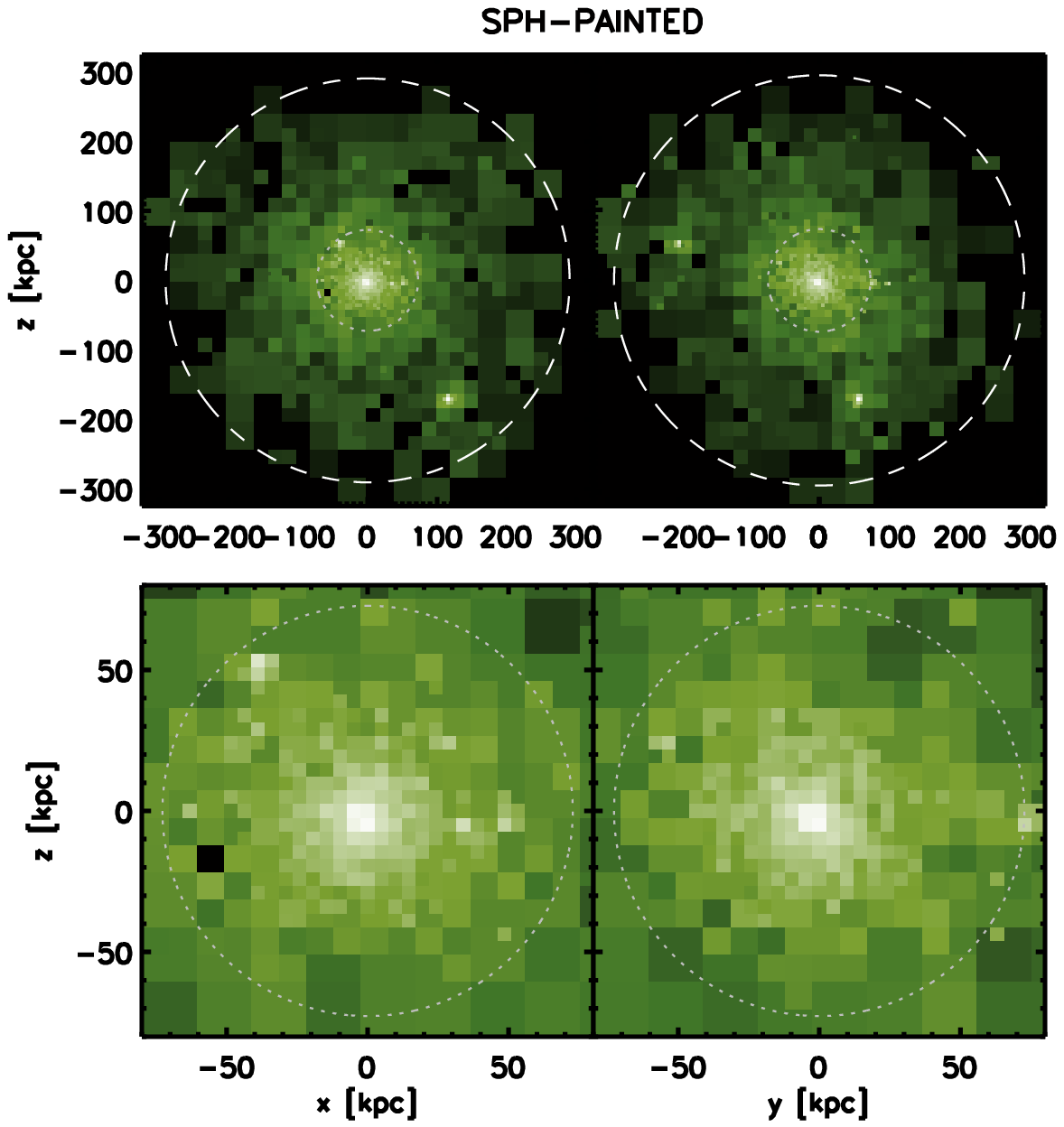} &
   \includegraphics[scale=0.65]{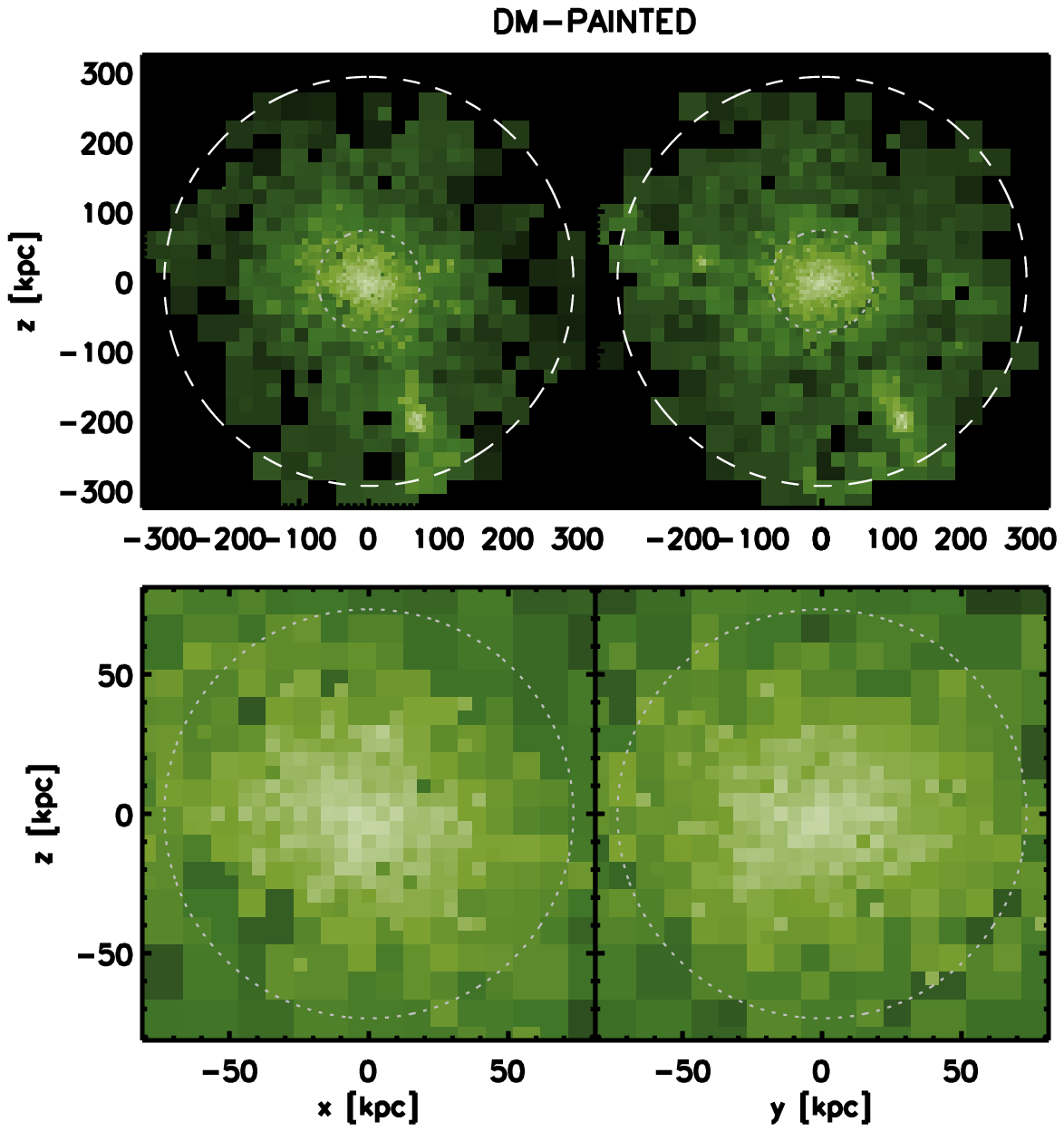}\\
 \end{array} &
\includegraphics[scale=0.4]{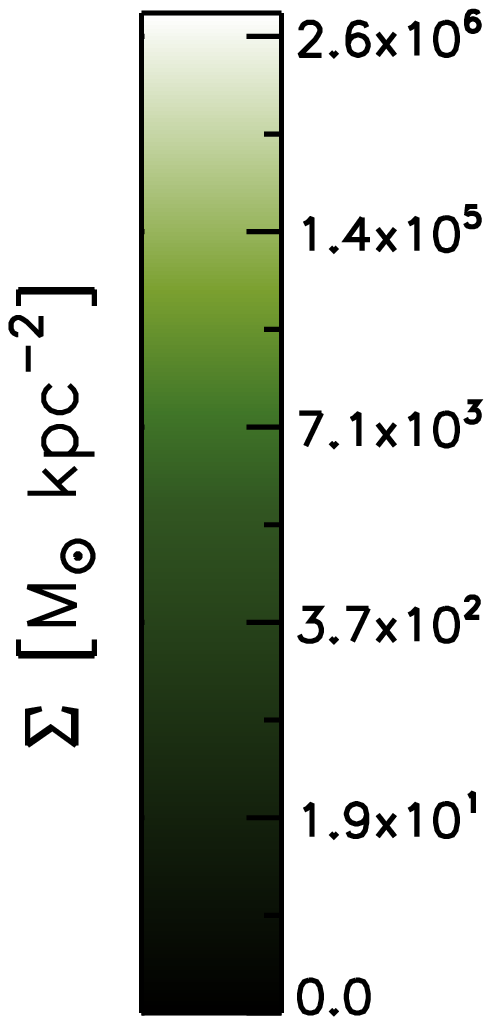}\\
\end{array}$
\caption{\label{fig:densmap}%
Projected stellar density map of modelled stellar halos in the MUGS g15784 simulation.
The top-left set of panels (SPH-STARS) shows all accreted stars in the SPH simulation; the top-right
set of panels (SPH-EXACT) shows DM particles in the SPH simulation painted with the same masses
as the self-consistently-formed stars; the bottom-left set of panels (SPH-PAINTED) shows the DM
particles in the SPH simulation painted according to the
best-fit stellar-DM mass relation;
and the bottom-right set of panels (DM-PAINTED) shows painted DM particles in the DM simulation.
Particles within $1.1~\rvir$ are plotted.
The pixel size is adaptively expanded from a minimum of $5$~kpc per side until
there are at least 5 particles per pixel, so the signal-to-noise per
pixel is approximately equal in the low density regions.
The top row of each set of panels shows the entire virial region, while the bottom row
is zoomed in by a factor of $4$. The gray dashed line denotes \rvir, and the gray dotted
line denotes $0.25~\rvir$.
The density scale is identical for all panels.}
\end{figure*}


\begin{figure*}
$\begin{array}{cc}
 \begin{array}{cc}
 \includegraphics[scale=0.65]{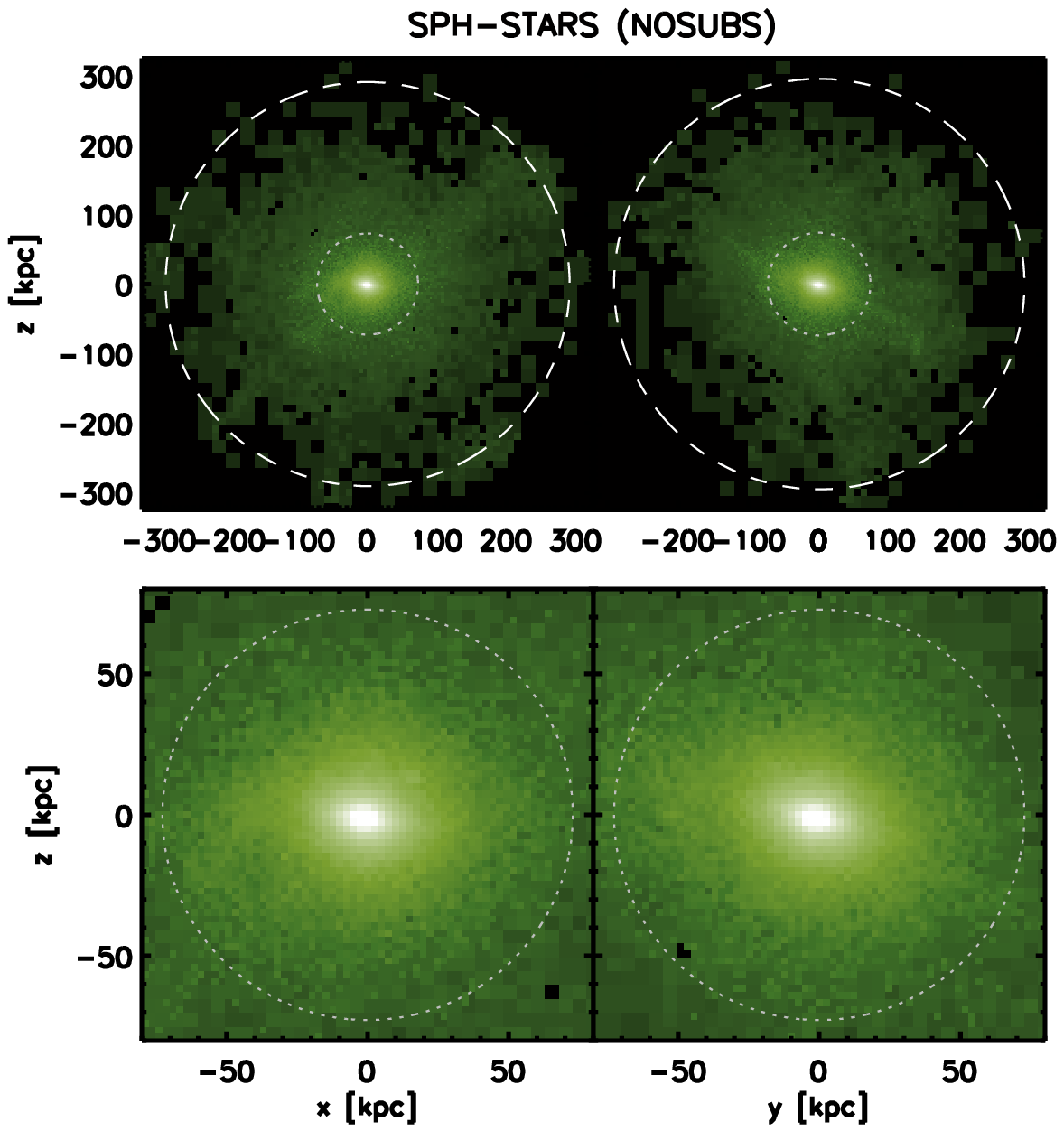} &
   \includegraphics[scale=0.65]{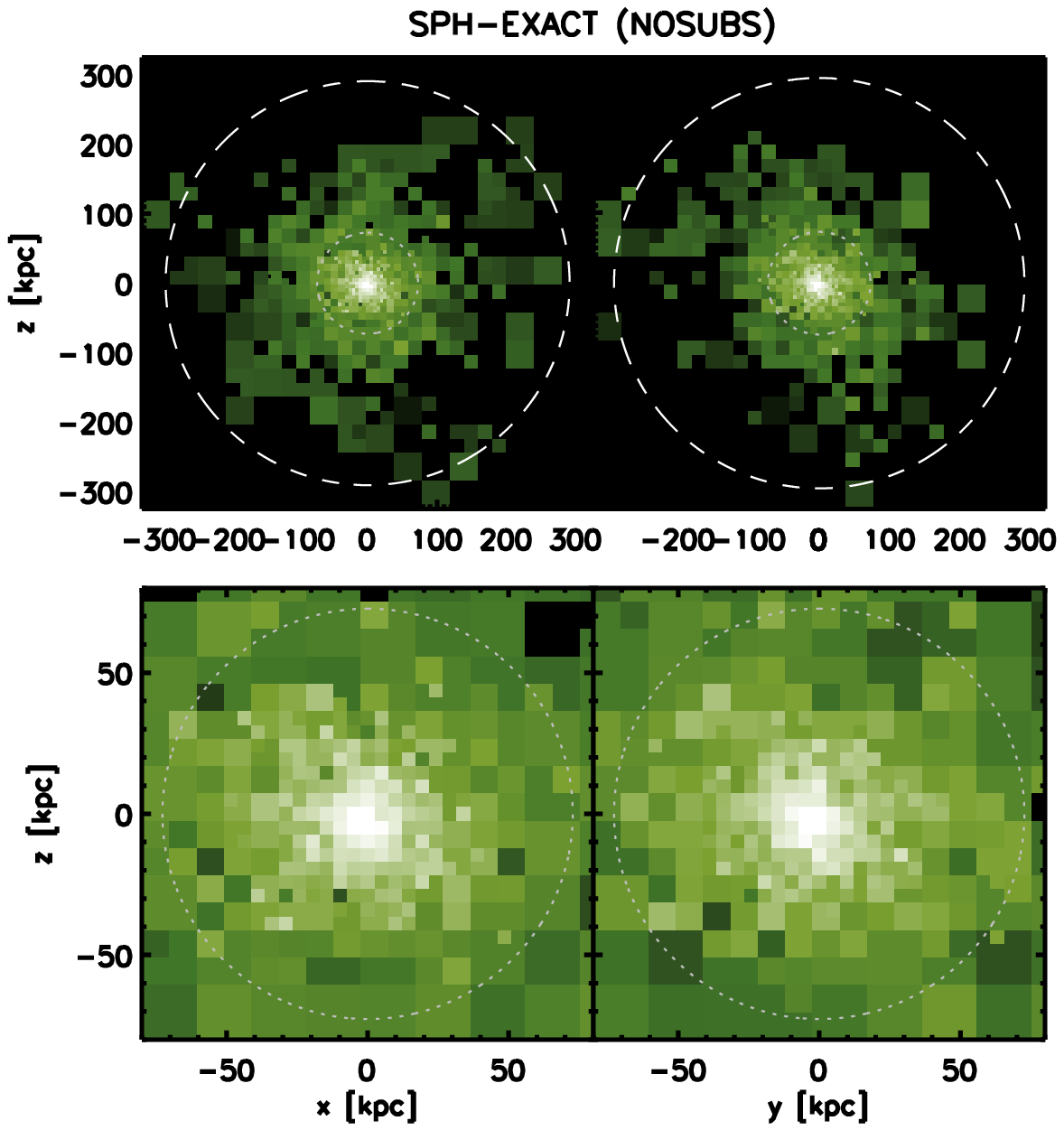} \\
 \vspace{0.5em}\\
 \includegraphics[scale=0.65]{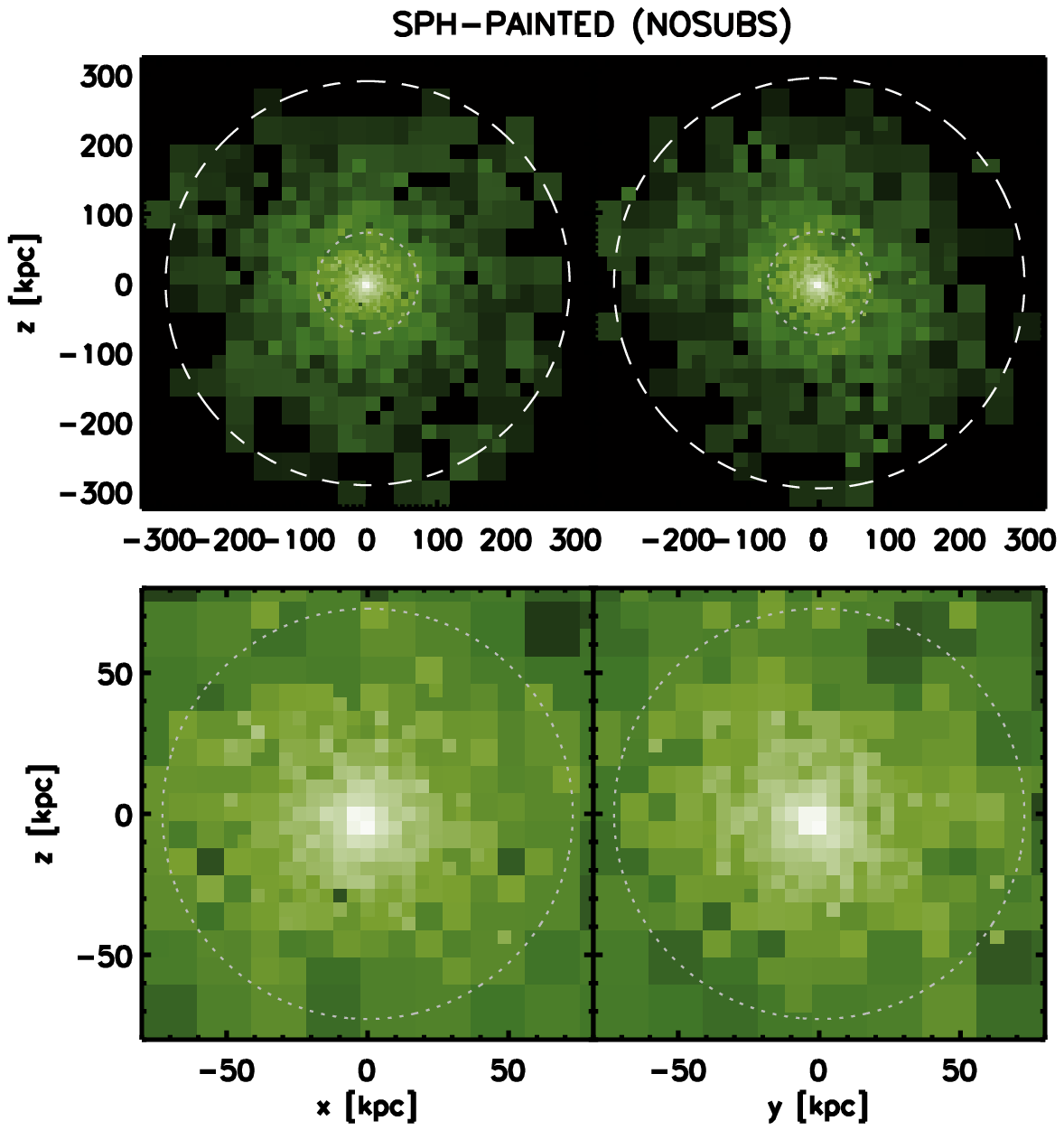} &
    \includegraphics[scale=0.65]{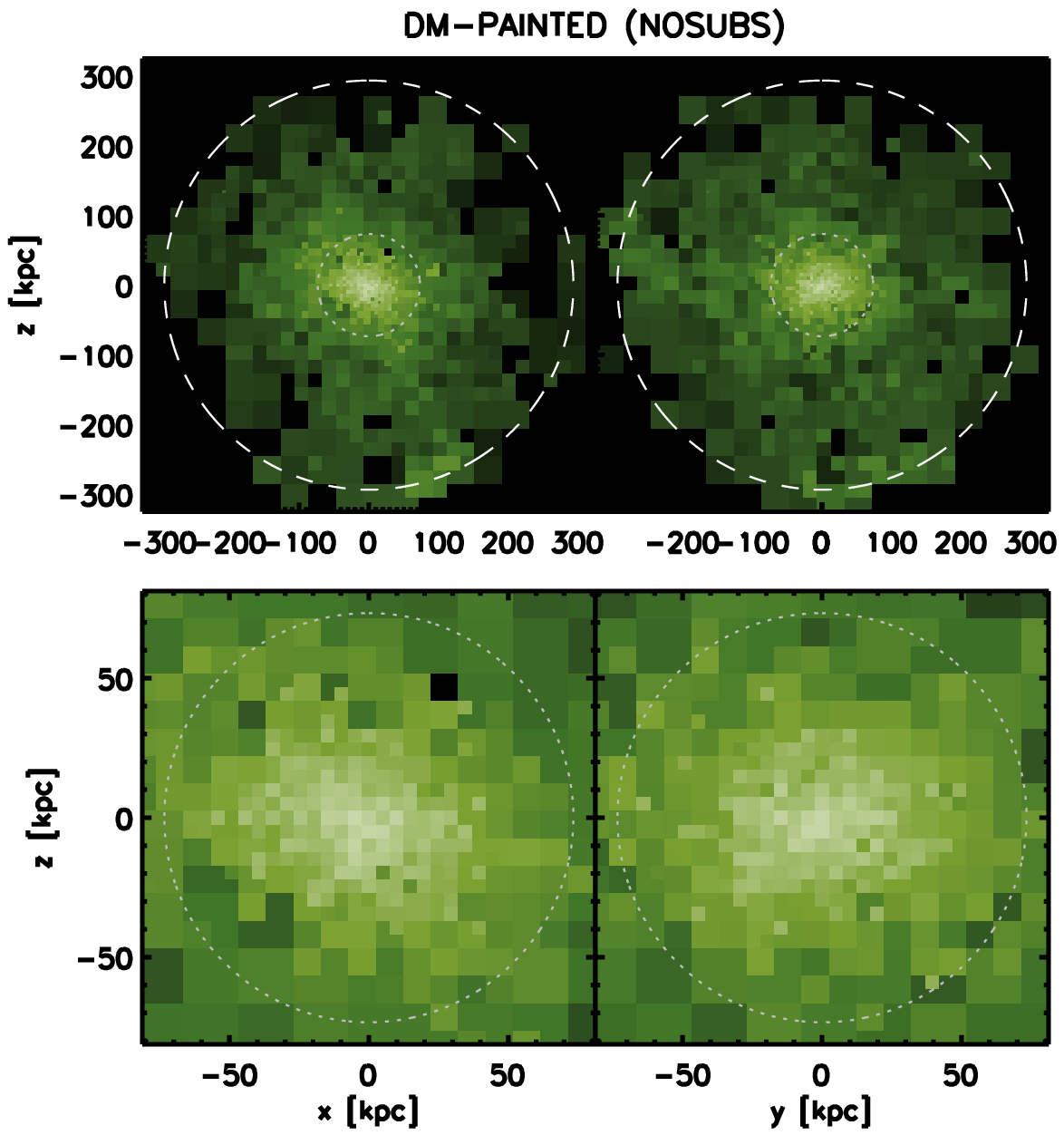} \\
 \end{array} &
\includegraphics[scale=0.4]{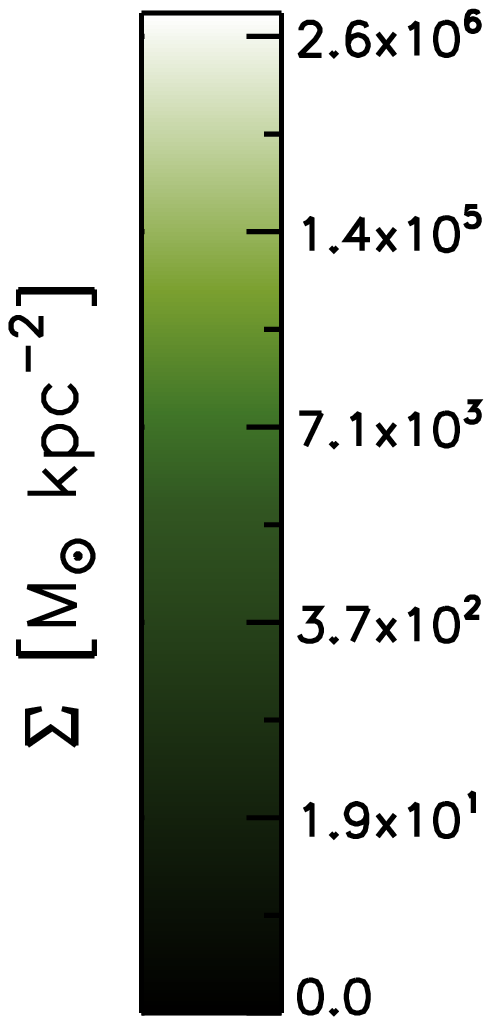}\\
\end{array}$
\caption{\label{fig:densmap_noss}%
As in Figure~\ref{fig:densmap}, but excluding bound subhalos.}
\end{figure*}

Two-dimensional projected maps of the density of the stellar halo models are shown in
Figure~\ref{fig:densmap}, which includes all accreted or painted particles,
and Figure~\ref{fig:densmap_noss}, which excludes those contained in bound substructure.
The pixel size is adaptively expanded from a minimum of $5$~kpc per side until there are
at least 5 particles per pixel, so the signal-to-noise per pixel is approximately equal
in the low density regions.

We first note that the models unambiguously trace the evolution of the same 
galaxy: the massive
satellites are recognizable in each model at similar locations. It is therefore valid to
directly compare the quantitative structure measurements in the different models and be confident
that the differences are due to differences in the assumptions of the models, not due to
different evolutionary histories.

Secondly, we note that the models show systematic differences. In particular, it is clear
that the global concentration and shape of the stellar halos are different. There also
appears to be more structure in the SPH-STARS model than in the SPH-PAINTED model, and yet
more structure to the DM-PAINTED halo. We will quantify these differences below,
which can be interpreted as an estimate of the magnitude of the systematic errors introduced
by the assumptions going into the halo models.

\subsection{Concentration}\label{sec:concentration}

\begin{figure}
\plotone{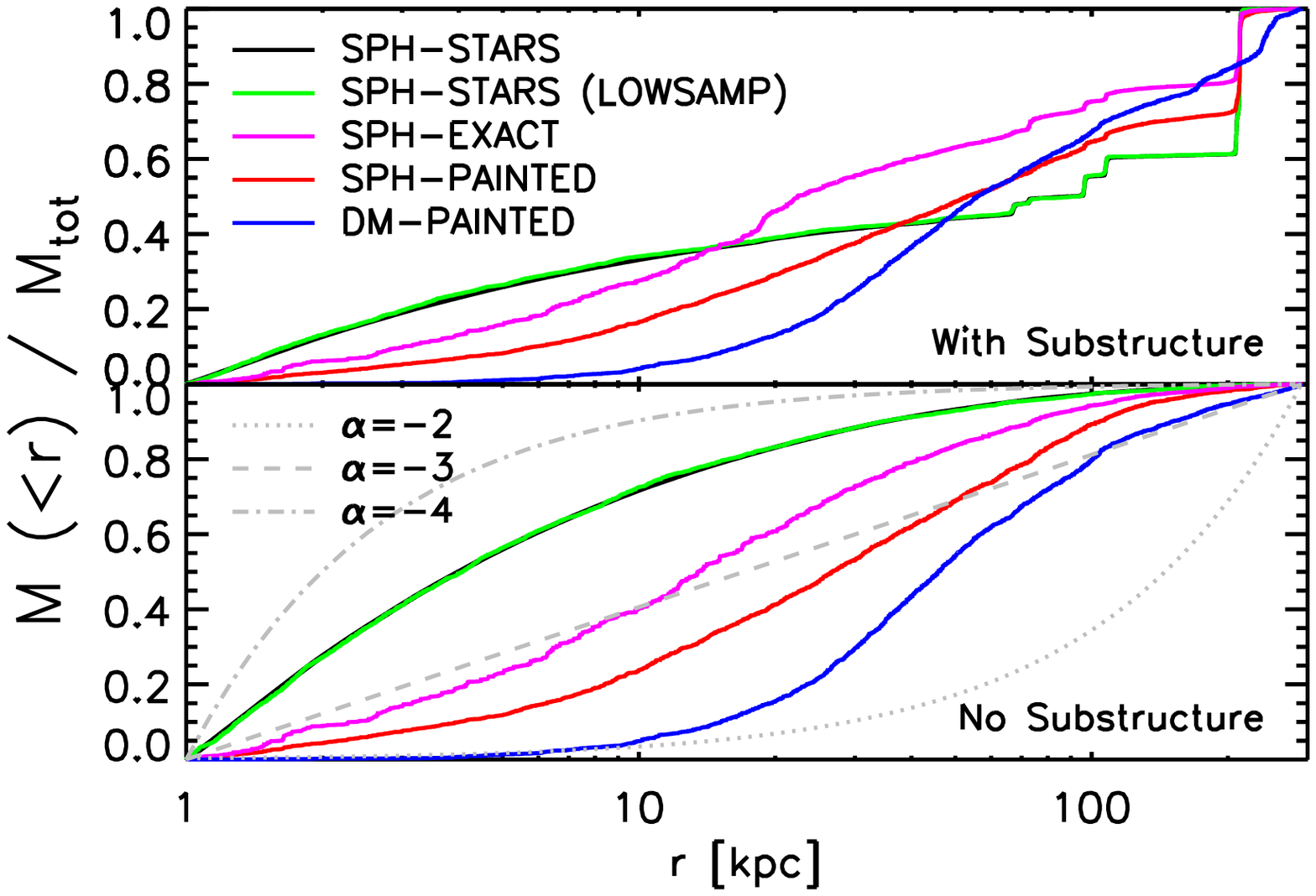}
\caption{\label{fig:masscprof}%
Cumulative radial profile of the stellar mass in the five stellar halo models, starting
at $1$~kpc. The top panel
contains substructure, while the substructure has been removed in the bottom
panel. The SPH-STARS model is significantly more radially concentrated than the SPH-PAINTED
model, which is itself more concentrated than the DM-PAINTED model.
The SPH-EXACT halo profile is similar to that of the SPH-PAINTED halo, but slightly more concentrated.
The smaller particle number in the LOWSAMP halo has no effect on the concentration, as it
lies essentially overtop the SPH-STARS halo.
The gray lines in the bottom panel indicate cumulative mass profiles
of halos with a power law density profile with slope $\alpha$ between $1$~kpc and the virial radius.}
\end{figure}

To measure the different concentrations of the stellar halo models, we plot their
cumulative stellar mass profiles in Figure~\ref{fig:masscprof}. In the top panel, we include
all star particles, while
we focus on the bottom panel, where the bound subhalos have been removed.
To assess the effect of painting, we compare the SPH-STARS and SPH-PAINTED models. The
model containing accreted stars is significantly more centrally concentrated than the painted
DM particles in the same simulation --- for example,
the half-mass radius is more than $6$ times smaller.
Because the SPH-STARS model has orders of magnitude more
particles, resolution could conceivably be an issue. To assess the impact of the particle
number, we have randomly sampled $5179$ particles out of the SPH-STARS model to form a new
``SPH-STARS (LOWSAMP)'' model that has the same number of particles as the SPH-PAINTED model;
this is shown in green, and has an identical radial distribution as the full SPH-STARS model.
We therefore conclude that the assumption that painted DM particles follow the same
distribution as self-consistently formed stars introduces a large systematic error in the
overall concentration of the halo. In the bottom panel, the cumulative mass profiles of
power law density distributions $\rho \propto r^{\alpha}$ have been overlaid for various
values of $\alpha$. Steep slopes of $\alpha < -3$ appear convex in this plot, while
shallow slopes of $\alpha > -3$ appear concave.
The SPH-STARS halo is well described by a power law with a slope of
$\alpha \sim -3.5$, while the SPH-PAINTED halo transitions from a relatively shallow $\alpha \sim -2$
in the inner regions to a much steeper $\alpha \sim -4$ slope in the outer region.

This general behavior agrees very well with what has been found in the literature.
\citetalias{bj05} find a halo density profile that transitions from
a shallow $\alpha \sim -1$ at small radius to $\alpha \sim -3.5$ at large radius,
similar to what is seen in the SPH-PAINTED model, which has very similar assumptions.
The accreted stars in \citep{font-etal11} have a power law slope of $\sim -3$ at most radii, steepening
to $-3.5$ at large radii, which is not dissimilar to what we find in the SPH-STARS model, and
while \citetalias{libeskind-etal11} never quantified the density profile
of the self-consistently-formed accreted stars in their hydrodynamic simulation, the cumulative
mass profile of such stars in their figure~1 is an excellent match to the analogous
SPH-STARS model in our Figure~\ref{fig:masscprof}.

If this difference is due to assigning an incorrect stellar mass to the accreted
halos, then the SPH-EXACT halo should mirror SPH-STARS; if it is due to the choice
of DM particles instead of stars, then it should mirror SPH-PAINTED.
Figure~\ref{fig:masscprof} shows that it is much more similar to SPH-PAINTED;
it has the same overall profile shape, and a much more similar half-mass
radius. However, it is undoubtedly more concentrated than SPH-PAINTED
(the half-mass radius is a factor of $2$ smaller).
This indicates
that halos that scatter high in Figure~\ref{fig:mstarcalib} end up at systematically
smaller radii, but that the more dominant effect is that painted DM particles are
less concentrated than accreted stars.

To assess the impact of the potential, we compare the SPH-PAINTED and DM-PAINTED models.
We find that they have similar functional forms, but that the SPH-PAINTED model is
significantly more concentrated; for example, the half-mass radius is $1.7$ times smaller.
Therefore, the baryonic contribution to the potential,
which is itself more centrally concentrated than the DM, leads to a more centrally concentrated
stellar halo.

\subsection{Shape}\label{sec:results-shape}

\begin{figure}
\plotone{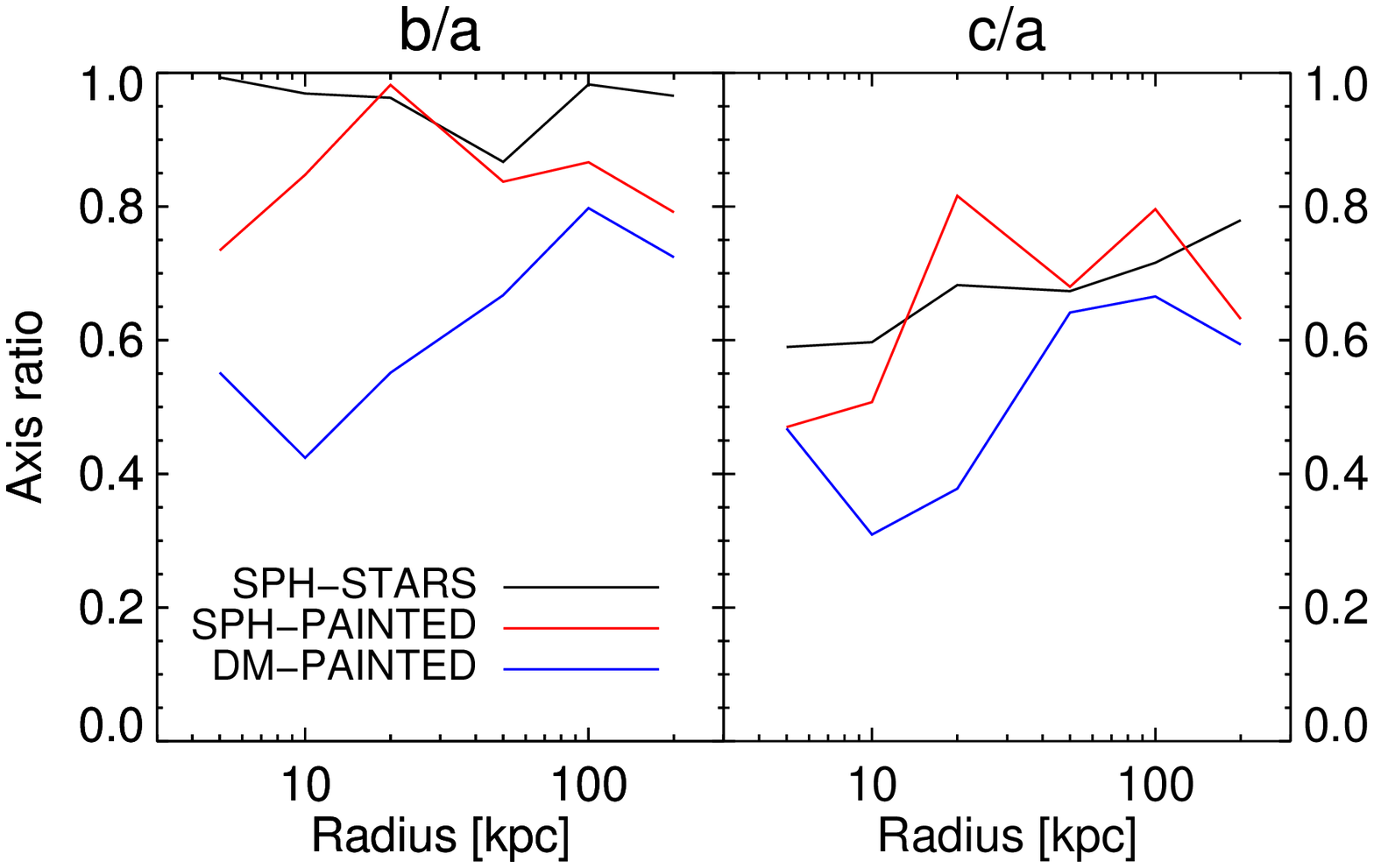}
\caption{\label{fig:axratplot}%
Intermediate \textit{(left)} and minor \textit{(right)} axis ratio of each stellar halo
model as a function of radius. Shapes are determined iteratively using the second moments
of the mass distribution within ellipsoidal shells of width $25\%$ of the radius, and
are plotted at the geometric mean radius of the principal axes. Bound substructures have been removed.}
\end{figure}

Another difference between the models is their global sphericity.
Figure~\ref{fig:axratplot} shows
the shape of the stellar distribution, which has been calculated using the 2nd moment tensor
of the stellar mass in an iteratively-defined ellipsoidal shell \citep[e.g.][]{zemp11}
of width $25\%$ of the quoted radius.
Both the SPH-STARS and SPH-PAINTED halos are somewhat oblate, with $b/a \sim 0.8$--$1$ and
a total flattening rising from $c/a \sim 0.5$ in the inner regions up to $0.8$ at the virial radius.
On the other hand, model DM-PAINTED,
which contains no disk, is very strongly prolate, with $b/a \approx c/a \sim 0.4$ -- $0.7$ depending
where it is measured.
This is not surprising,
since the dark matter halos of simulations with disks are strongly modified by the
presence of the disk, becoming less flattened and more oblate, relative to the more flattened
prolate dark matter halos that predominate in pure DM cosmological simulations
\citep[e.g.][]{kazantzidis-etal04,bailin-etal05-diskhalo}.

\subsection{Substructure}

A key prediction of stellar halo models is the degree of substructure. A rough
measurement of this is the variation in the stellar mass density within a
shell of a given
radius; this is similar to the ``sigma/total'' measurement used by
\citet{bell-etal08}. Formally, we divide the virial region of the halo into
(initially) spherical shells, and then subdivide each shell
into angular sectors of equal volume. The divisions between these sectors
are spaced equally in azimuthal angle $\phi$ and in the cosine of
the polar angle $\theta$. We use $N_{\phi}=4$ azimuthal divisions and
$N_{\theta}=4$ polar divisions. This probes different physical scales at different
radii, and therefore one should not compare the quantitative measurements between
radial bins, but rather compare different models at the same radius.
We compute the mean stellar mass density of each model within the entire
shell \meanrho, and the root mean square (rms) of the sector-to-sector
variation, $\sigma_{\rho}$. There is some contribution due purely to shot noise
from the finite number of particles, $\sigma_{\mathrm{shot}}$, the magnitude of
which can be determined by noting that the total mass $M$ within a sector
is equal to the number of particles $N$ times their mean mass \meanm:
\begin{equation}  M = N \meanm  \end{equation}
\begin{equation}
 \sigma^2_{\mathrm{shot}} = \left(\frac{\df M}{\df N}\right)^2 \sigma^2_N + 
  \left(\frac{\df M}{\df\meanm}\right)^2 \sigma^2_{\meanm}
\end{equation}
\begin{equation} = \meanm^2 N + N^2 \left(\frac{\sigma^2_m}{N}\right) \end{equation}
\begin{equation} = \meanm^2 N + N \left( \left<m^2\right> - \meanm^2 \right) \end{equation}
\begin{equation} = N \left<m^2\right>. \end{equation}
Technically, this derivation assumes that the mass per sector within each radial
bin is independent, while in reality there is an additional constraint that the sum
of the masses of the sectors must equal the mass in the shell. However, with $16$
sectors, the reduction of one degree of freedom only changes the shot noise by $\approx 3\%$.
We have verified the accuracy of this analytic expression using Monte Carlo experiments.
The intrinsic sector-to-sector dispersion $\sigma_{\rho}$ is then the measured
rms minus the shot noise $\sigma_{\mathrm{shot}}$, in quadrature.

An additional complication is that the shapes of the halos are different.
As noted by \citet{knebe06}, densities at a given radius can vary by $10$--$50\%$
due to the ellipticity of the density distribution, which could dominate the
sector-to-sector dispersion if not taken into account. We therefore calculate
the shape of the density distribution in a shell of geometric mean
radius $30$~kpc and width $20$~kpc, using the method described
in Section~\ref{sec:results-shape},
and use these principal axes to define ellipsoidal shells in which to
determine $\sigma_{\rho}$. These shells have the same width and geometric
mean radius as the corresponding spherical shells, and therefore
the same volume.
We also scale the particle coordinates in the principal axis frame by
the lengths of the principal axes before we determine which angular
sector it belongs to; this ensures that the
sectors all have equal volumes regardless of the shape of the ellipsoid.

\begin{figure*}
\plottwo{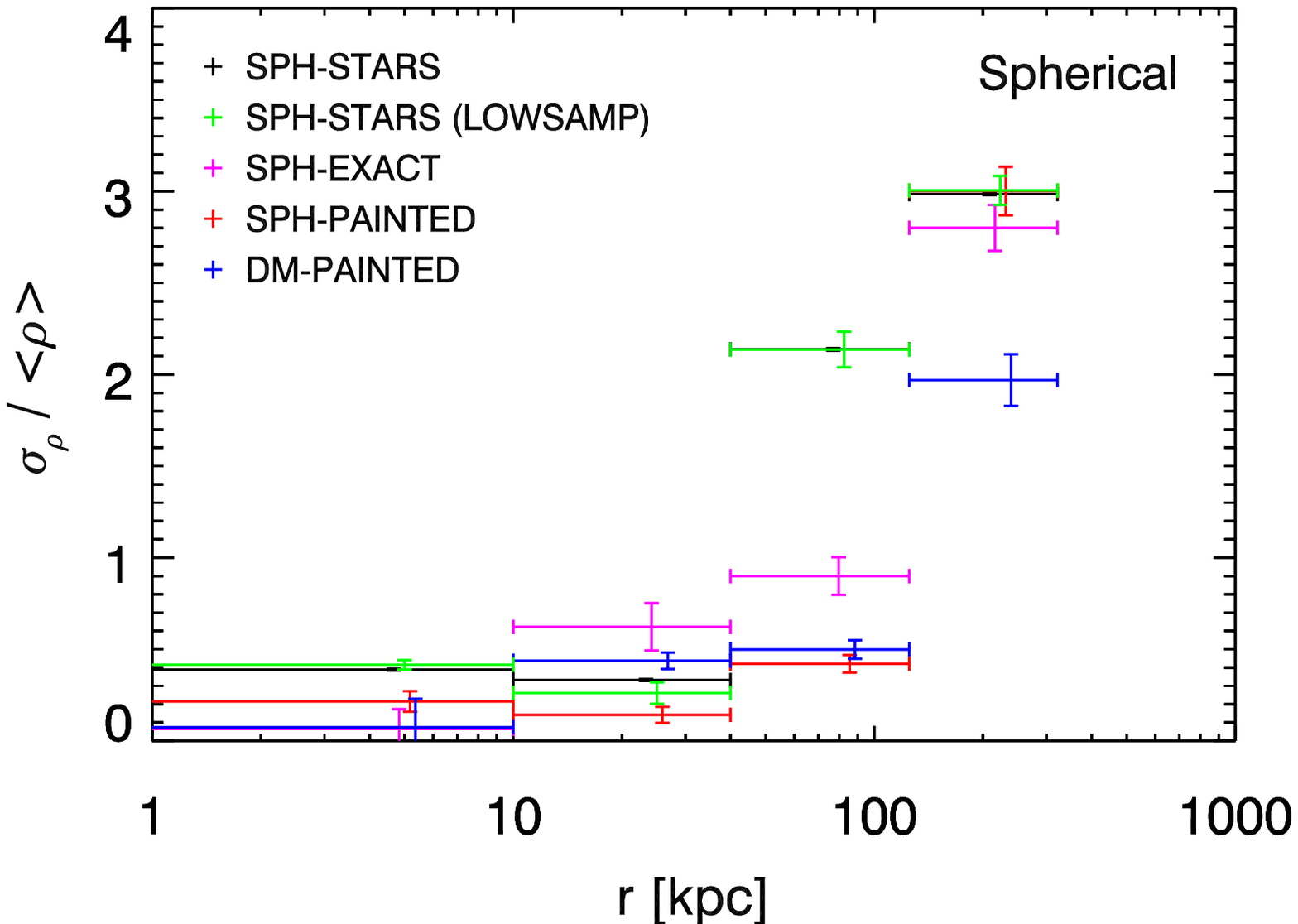}{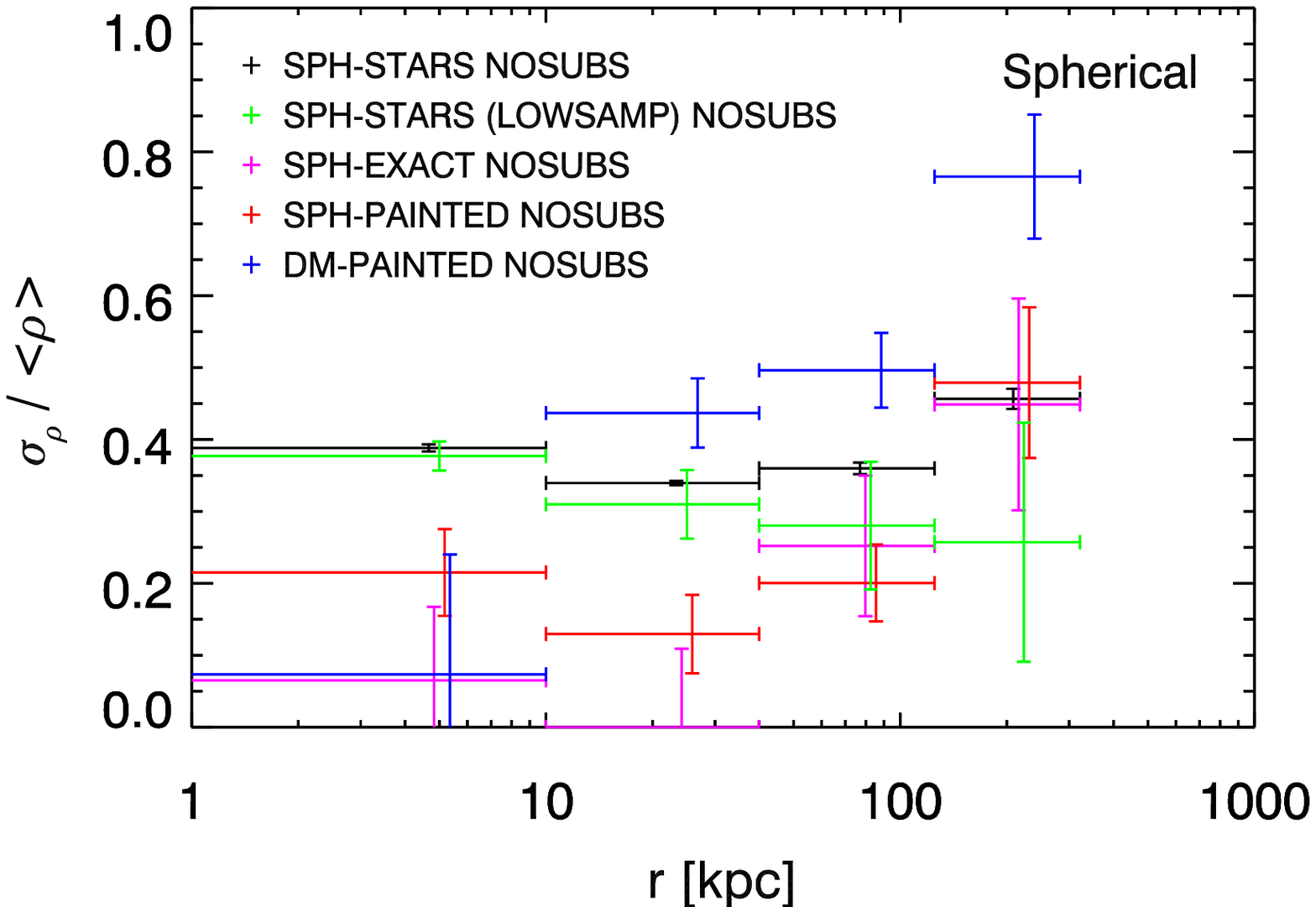}\\\vspace{1em}
\plottwo{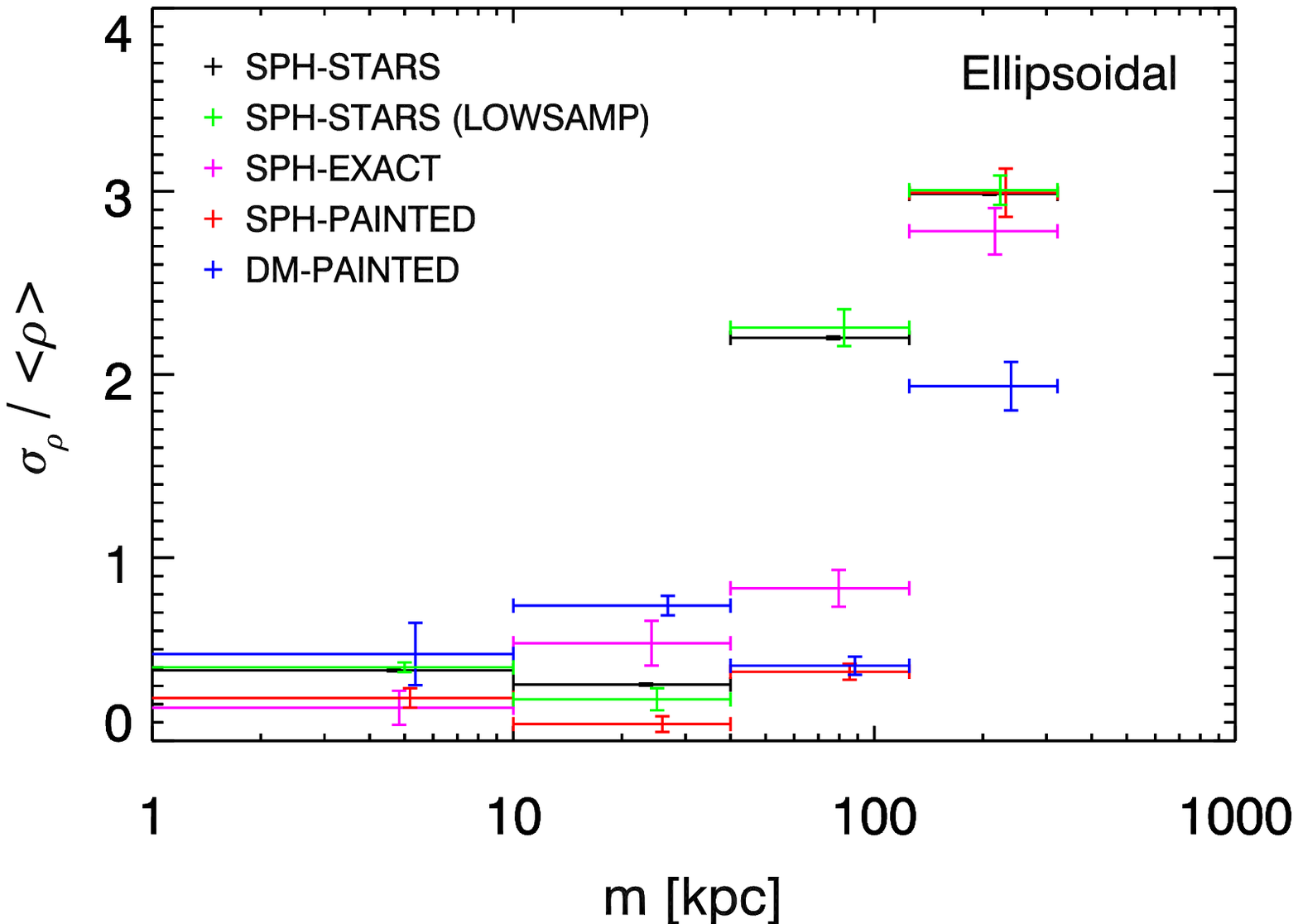}{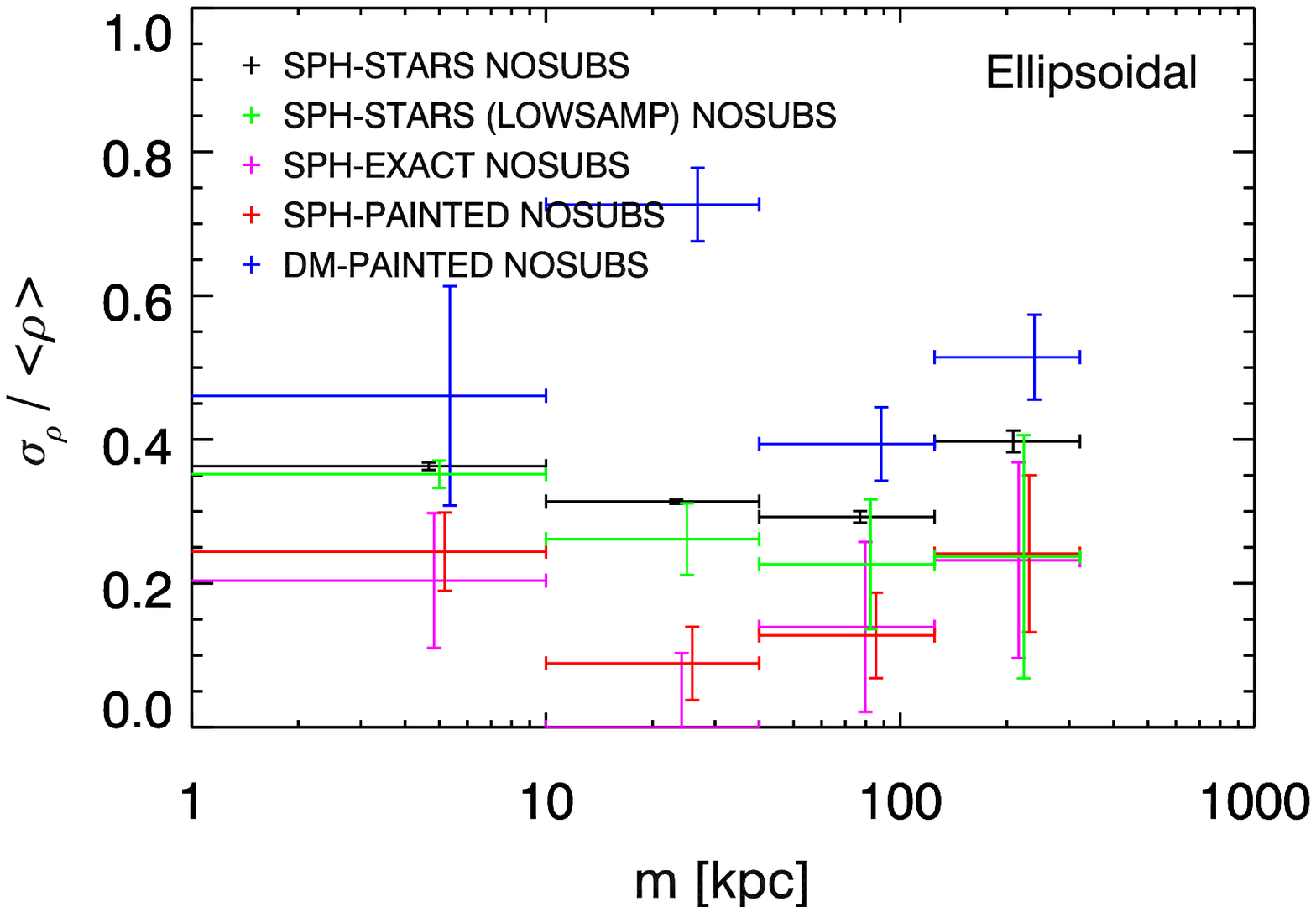}
\caption{\label{fig:sigma}%
Sector-to-sector dispersion $\sigma_{\rho}$ of the stellar mass density within spherical (top panels)
and ellipsoidal (bottom panels) shells, relative to the mean density $\meanrho$ within the shell.
Shot noise due to the finite number of particles has been subtracted in quadrature.
The left column indicates the results for all relevant particles, while the right column
excludes bound subhalos.
The preferred panel is on the bottom-right (Ellipsoidal NOSUBS).
There are $N_{\theta}=4$ polar sector divisions and $N_{\phi}=4$ azimuthal sector divisions.
Horizontal error bars indicate the radial extent of each bin, while the vertical error bars
are the bootstrap error bars in $\sigma_{\rho}/\meanrho$.}
\end{figure*}

The amount of structure seen in each halo model is shown in Figure~\ref{fig:sigma}.
The lefthand plots include all stellar mass, while the bound substructures have been removed
in the righthand plots. The top plots show the dispersion within spherical shells, while
the bottom plots use the ellipsoidal shells discussed above.
The lefthand plots make it
apparent that the different models sometimes predict dramatically different amounts
of substructure at the same radius;
however, there is no clear systematic pattern to the differences.
Much of the substructure in these plots is due to the distinct satellite galaxies in the
simulation \citep[see][]{nickerson11} rather than the diffuse halo,
which are usually excluded from observational studies \citep[e.g.][]{bell-etal08}.
We therefore focus on the righthand plots, in which bound substructures have been removed,
and particularly on the bottom-right panel, where the ellipsoidal shape
of the halo has been taken into account.

The number of particles could impact the amount of structure, and
certainly the error bars, complicating the comparison between the SPH-STARS and
SPH-PAINTED halos. We therefore first compare the SPH-STARS (black) and SPH-STARS (LOWSAMP)
(green) halos and find that the error bars are indeed significantly larger at
lower resolution, but that the results always agree to within the error bars,
giving us confidence that the error bars accurately portray the uncertainty in
the measurement. Although it appears that the subsampled halo is systematically
less structured, this is a coincidence of the randomly-sampled particles that
constitute the LOWSAMP halo; with different random seeds, the LOWSAMP halo scatters
both positive and negative around the full resolution halo, with a dispersion
comparable to the quoted error bar. We therefore conclude that differences between
halo models that are larger than the error bars are real discrepancies between
the predictions.

We assess the impact of using painted dark matter particles instead of stars
by comparing SPH-STARS (black) to SPH-PAINTED (red).
SPH-PAINTED is systematically
less structured at all radii when using our preferred ellipsoidal bins,
typically by a factor of $2$. There is little difference between the dispersion
measured in spherical vs. ellipsoidal shells, which is consistent with the
similar global shapes of these halos (Figure~\ref{fig:axratplot}).
The SPH-EXACT halo (magenta) is indistinguishable from the SPH-PAINTED halo,
indicating that the difference in the amount of substructure is entirely due to the use of painted DM
particles.

The impact of the different potential in the baryonic simulations can be seen
by comparing SPH-PAINTED (red) to DM-PAINTED (blue). The dark matter-only simulation
is systematically more structured at all radii, in this case by factors of typically
$3$, although as high as $7$ at intermediate radii. Although the global halo
shapes are different enough that they could introduce a significant dispersion
in a structureless halo, the same discrepancy is seen when using spherical
shells at almost all radii, allowing us to be confident that this is not
an artifact of the more flattened halo in the dark matter-only simulation.

\section{Discussion}

We have tested the effects of two common assumptions used in stellar halo
models: (1) that stellar mass can be painted onto DM particles, and
(2) that the baryonic changes in the potential can be ignored. We discuss
below the effects of each assumption.

\subsection{Painting}
When we compare the SPH-STARS and SPH-PAINTED halo models, we find that
SPH-PAINTED is less concentrated and less structured than SPH-STARS.
Both models put, on average, the same stellar mass into the same progenitor
objects and evolve them in the same potential, so it is surprising that
there is such a large difference. This difference is entirely due to
(a) assuming that the scatter in the stellar-DM mass relation does not
correlate with the fate of the accreted halo, and
(b) assuming that DM particles deep in the potential well of a subhalo
(note that we only paint the most bound $1\%$ of the DM particles)
evolve similarly to the star particles, which are also found deep
in the potential well.
The SPH-EXACT halo, where the scatter is not an issue, is
slightly more concentrated than SPH-PAINTED, indicating there
is a slight tendency for halos with high stellar masses to wind up
in the inner parts of the halo, but that this effect does not
dominate the overall radial profile.
Moreover, SPH-EXACT and
SPH-PAINTED have identical substructure, indicating that the
main reason that the SPH-STARS and SPH-PAINTED models differ is
because DM particles deep in the potential well are different
from stars deep in the potential well.
Although these particles are co-located, the kinematics of the populations are
not identical: the DM particles are dynamically hotter, having
orbits that take them further
from the center of the subhalo (a consequence of the more
extended nature of the DM component).
We postulate that this difference causes the DM particles to
be stripped earlier than stars during the subhalo's accretion and
orbit around the parent galaxy, and results in a less concentrated halo.
The higher velocity dispersion of the DM also means the stripped debris is less
coherent, resulting in less substructure.
In both cases the debris orbits in the same potential, and it is therefore
not surprising that painting does not affect the shape of the predicted
stellar halo.

It is important to note that we have formally only tested one particular
painting scheme, and this scheme is less sophisticated than many that are used.
Therefore, we must be careful about what lessons are generalizable.

Firstly, we note that it is a generic property of galaxy formation physics
that baryons are more concentrated than DM within subhalos, and as a
direct consequence the DM particles at the same radii as star particles
have different kinematic properties such as velocity dispersions and angular momenta.
We therefore expect these results to generalize
to any painting scheme that does not explicitly guarantee that the painted
DM particles share not only the same spatial distribution as the
stars within satellite galaxies, but also the same kinematic distribution.

Secondly, it appears that the systematic difference that painting induces
in the halo concentration
can completely explain discrepancies in the literature between
the radial density profiles in the models of \citetalias{bj05}, which used
painting, and the accreted stellar halo in the hydro simulations of
\citetalias{libeskind-etal11}, which did not.

\citetalias{libeskind-etal11} advocate one particular painting scheme where DM
particles are chosen to lie within a given depth of the subhalo potential
well, and demonstrate that the resulting painted stellar halo model has the
same concentration as the accreted stars in their hydro simulation.
This painting scheme requires higher
resolution than present in our simulations, so we cannot directly test it,
but we find the match between the concentration of the stars and
painted DM particles very encouraging;
testing to see if the painted DM halo has the same degree of substructure
as the star particles is another critical test that we would strongly advocate.
However, there is one important aspect of this scheme that may compromise its use
for stellar halo models: the scheme was calibrated so that the radial
distribution of the diffuse halo, i.e. the stripped satellites, matched
the SPH stars\footnote{Note that although the scheme was calibrated to recover
the correct radial distribution, the fact that it was able to do so via
tuning one parameter is not a trivial result: as we have demonstrated, many other
painting schemes are not able to do this for \textit{any} choice of parameters.}
without regards to the properties of the satellites themselves.
There is therefore no guarantee that the stellar masses
(and consequently metallicities; \citealp{tremonti-etal04}) or radial profiles
of the satellites are correct, and in fact, scaling arguments suggest
that the stellar masses, in particular, may not scale correctly
with total DM halo mass. In other words, it may be that
\citetalias{libeskind-etal11} are building a realistic-looking halo out of
the wrong pieces; if this is true, then properties like the metallicity structure,
which is one of the key observables one would like to extract from stellar
halo models \citep[e.g.][]{font-etal06}, will be incorrect.
Reproducing the properties of the satellites themselves is a critical
test that any painting scheme must pass, and it is not immediately obvious
whether it is possible for \textit{any} method of painting DM particles to satisfy
all of the necessary constraints.

The most important and general point is that painting 
can easily introduce systematics at the factor-of-several level.
We strongly urge modelers to perform tests like these
to estimate the magnitude of the systematic error when introducing new
painting schemes.

\subsection{Baryonic Potential}\label{sec:discussion-potential}
When comparing the SPH-PAINTED and DM-PAINTED halo models, we find that
DM-PAINTED is less concentrated, its shape is more prolate and less flattened,
and it has more substructure. Both models use the same painting scheme on
the same progenitors, but evolve them in a different gravitational potential.
The critical differences between the potentials are that the SPH-PAINTED
model has a parent galaxy that is more concentrated and has a disk,
and also has satellites with deeper potential wells.

The global shape of the potential clearly has an impact on the global
shape of the stellar halo: when the potential is more concentrated,
so is the stellar halo; when the potential is more spherical,
so is the stellar halo; when the potential is more prolate,
so is the stellar halo.
The baryons also make it more difficult to strip particles out
of the satellites, so the satellites must get closer to the
center to get stripped and therefore deposit their tidal debris
closer to the center.

The reason for the different amounts of
substructure is less clear. There are two physical mechanisms that
could decrease the amount of substructure in SPH-PAINTED:
differential precession of streams in the oblate potential of the disk,
and changes in individual orbits due to the central concentration
of the potential (below we will refer to this as ``scattering'' for
simplicity, although for an extended central concentration, like
a disk, this is primarily due to individual orbital types changing
their shape rather than true scattering onto chaotic orbits that
is seen for point-like central concentrations; \citealp{debattista08,valluri10}).
We do not have the ability to independently assess each effect
with this simulation, but we note that the global potential in
the DM-only simulation is more prolate-triaxial than in the SPH simulation,
and it is only in the disk region that the SPH potential has a significantly
smaller $c/a$ axis ratio.
We may therefore expect differential precession to disrupt
streams in the outer parts of the halo faster in the DM-PAINTED model but
streams in the inner parts of the halo faster in the SPH-PAINTED model.
Instead, the SPH-PAINTED model has less structure at all radii, but the
effect is indeed strongest within $40$~kpc. This suggests that both
differential precession and orbital scattering are playing a role.

The overproduction of substructure in pure DM models must be taken
into account when comparing these models to observations. For example,
\citet{helmi-etal11} determine that the Milky Way halo contains less
structure than predicted by the \citetalias{cooper-etal10}
pure DM models,
and conclude that the Milky Way halo must also contain a smooth in situ
component that reduces the total substructure. However, the factor
by which \citet{helmi-etal11} find that the 
model overpredicts the structure is $2$--$3$,
which is of the same magnitude as the level of systematic
overprediction of substructure we find for pure DM models.
We therefore urge caution against overinterpreting differences
between these models and observations that are smaller than
the scale of the systematics that we find.

\subsection{Simulation Caveats}
Because the galaxy in the SPH simulation is not a perfect representation of
a real galaxy, it is worth discussing how those differences might affect
our conclusions.

The primary differences between the simulated galaxy and a typical disk galaxy
of the same mass are that (1) its potential is too concentrated and (2) its
disk fraction is too small.
We therefore might expect that effects we
see that are due to the concentration of the potential might
be overestimated, while those due to the diskiness of the potential might
be underestimated.

We argue that the systematic error due to painting primarily arises because the kinematics of
DM particles deep in the potential of a satellite are different from those of stars deep
in the potential. In a more realistic less concentrated potential, the kinematic differences
between different
particle types might be expected to be less,
suggesting an overestimate of
the level of this systematic effect. On the other hand, if the stellar distribution
were diskier, the kinematics of the stars in that rotating disk would differ even more
from those of the non-disky DM particles, suggesting we are underestimating the level
of this systematic.
We do not know a priori which of these effects would dominate,
so the quantitative degree of the effect is uncertain, but the overall sign and
approximate magnitude of the systematic are likely to be faithfully indicated
by this work.

The SPH-PAINTED stellar halo is more concentrated than the DM-PAINTED stellar halo
due to the concentration of the baryonic potential. Because the baryonic potential is
too concentrated in the SPH simulation, the degree to which the baryons concentrate
the stellar halo could also be overestimated.

The amount of substructure in a more realistic potential would be expected
both to increase because of the less concentrated potential
and decrease because of the diskier potential --- we believe both phenomena are
important in determining the amount of substructure, as discussed
in section~\ref{sec:discussion-potential}.
Again, this may affect the quantitative measurement but is unlikely to
affect the sign or overall magnitude of the systematic effect,
which are our main conclusions.

One further caveat is that the dynamical mass of star particles in the SPH-STARS model is
significantly smaller than the mass of the dark matter particles,
so the central region of the SPH simulation, where the stars dominate, is effectively
simulated at higher resolution than in the DM-only simulation.
One could therefore imagine that some of the differences we see are related
to resolution rather than to baryonic effects. However, it is unlikely that this
would drive our results. Increasing the mass resolution of a DM-only simulation, as we would
need to do to match it to the mass resolution of the SPH stars, does not result
in different structure at the scales of interest: higher-resolution DM halos are
comparably triaxial, have comparable radial profiles, and have comparable numbers of
higher-mass (i.e. resolved in the lower-resolution simulation) subhalos
\citep[e.g.][]{stadel09,gao11}.
A related issue, however,
is that hydrodynamic simulations are much more sensitive to resolution, and so the detailed
structure of our SPH galaxy and therefore of the stellar halo models built from it
could be different than that of a galaxy simulated at higher resolution. However,
the main reasons that we see a difference are because the baryons form a concentrated distribution,
and because that distribution is disk-shaped; both of these facts will be true
of any hydrodynamic simulation that forms a disk galaxy. Therefore, while the detailed
structure of the halo model may differ at different resolutions, the comparison between
the halo models should still yield a good estimate for the magnitude of the systematic
effects that are introduced by the modeling assumptions.

Overall, we therefore caution that the differences between the simulated galaxy
and real galaxies may have a quantitative effect on our results.
However, the sign and approximate magnitude of the systematic differences
reported here are expected to be robust, with the possible exception of the
large differences in concentrations between the SPH halo and the dark matter
only case, which is expected to persist qualitatively but quantitatively may
be substantially overestimated in a simulation such as ours with an overly
concentrated baryonic component.

\section{Conclusions}
We have examined two critical assumptions that are part of most models
of the structure of stellar halos: using painted DM particles to represent stars, and
the omission of baryonic contributions to the gravitational potential.
We have used a controlled set of models where we can independently test their
effects. In one test, we compare stars formed in a cosmological hydro
simulation to painted DM particles in the same simulation to test
the effect of painting, while in the
other test we compare DM particles painted in a hydro simulation
to DM particles painted in the identical way in a pure N-body realization
of the same initial conditions
to test the effect of the different potentials.

We find that both sets of assumptions cause significant differences in the properties
of the predicted stellar halos. Painting results in a \textit{less concentrated}
halo, with a half-mass radius $\sim 6$ times larger, and a
\textit{less structured} halo, by a factor of $\sim 2$. Some of the
concentration difference is due to a systematic tendency for progenitors
with high stellar mass to wind up at small radius,
but most of the concentration difference and all
of the structural difference is due to the different kinematics of DM particles
and stars at the same location within a satellite.
The omission of the
baryonic contribution to the potential results in a halo that is
\textit{less concentrated}, by a factor of $~1.7$ in half-mass radius,
\textit{more structured}, by a factor of $2$--$7$,
and \textit{more prolate}, with $b/a \sim c/a \sim 0.6$.
The mechanisms that drive these changes are likely a combination of
orbit scattering from the central density enhancement, differential precession
when orbits are near the disk, and the overall prolateness of the
dark matter halo.

This is the first attempt we are aware of to
estimate the magnitude of the systematic effects present in stellar halo models
based on dark matter simulations,
and the results are somewhat discouraging.
The factor-of-a-few level of systematic uncertainty is similar to the
factor by which some of the models are discrepant from observations of the Milky
Way halo, meaning we cannot
presently conclude anything about the origin of the halo
from that discrepancy.

The hydrodynamic simulations differ from observed galaxies in some important ways
that could affect our results; most importantly, the baryons in the simulation are more
concentrated, and the baryonic distribution is less disky. These could affect
the quantitative measurements that we make, but because these two differences
act in opposite directions, the overall sign and approximate magnitude of the systematic
error we measure is unlikely to be affected (with the possible exception of the
halo concentration, which could potentially be substantially overestimated).
We also caution that parts of the simulations we are comparing operate at different
resolution due to the presence of low-mass star particles; however, we have
argued that this is unlikely to undermine our conclusions.

Are there potentially methods that can be used to create high resolution stellar
halo models that circumvent these difficulties?
One possible way forward, which we are pursuing,
is to couple a very sophisticated painting technique
to an evolving halo potential that both self-consistently solves the DM
dynamics and includes an analytic disk (which should itself be consistent
with the properties of a galaxy forming within that DM halo, for example
via a semi-analytic model).
We stress that it is important for any painting technique to be tested, using methods
such as the one we have used in this paper, to ensure that it reproduces
the spatial distribution and substructure expected from stars that form
with the same efficiency,
and that it reproduces the properties (such as stellar mass fraction) of the
satellite galaxies. Such a hybrid approach is likely to provide the
best hope for modeling stellar halos in the near term.
Looking forward, the highest resolution hydro simulations today are beginning
to be able to produce interesting predictions, and future hydro simulations with hundreds
of millions of particles within the virial radius are likely to provide the best
models once they are feasible.

\acknowledgments
We thank the referee, Joop Schaye, for a helpful report.
This work has made use of the Shared Hierarchical Academic Research Computing
Network (SHARCNET) Dedicated Resource Project: `MUGS: The McMaster Unbiased
Galaxy Simulations Project' (DR316, DR401 and DR437).
This work was supported in part by National Science Foundation
grant AST 1008324.
MV was supported by National Science Foundation grant AST-0908346.
VPD is supported by STFC Consolidated grant \# ST/J001341/1.
GS received funding from the European Research Council under
the European Union's Seventh Framework Programme (FP 7)
ERC Grant Agreement n. [321035].
HMPC and JW acknowledge financial support from the National
Science and Engineering Research Council.

\bibliography{/Users/jbailin/Dropbox/masterref}

\begin{thebibliography}{38}
\expandafter\ifx\csname natexlab\endcsname\relax\def\natexlab#1{#1}\fi

\bibitem[{{Bailin} {et~al.}(2011){Bailin}, {Bell}, {Chappell}, {Radburn-Smith},
  \& {de Jong}}]{bailin11}
{Bailin}, J., {Bell}, E.~F., {Chappell}, S.~N., {Radburn-Smith}, D.~J., \& {de
  Jong}, R.~S. 2011, \apj, 736, 24

\bibitem[{{Bailin} {et~al.}(2005){Bailin}, {Kawata}, {Gibson}, {Steinmetz},
  {Navarro}, {Brook}, {Gill}, {Ibata}, {Knebe}, {Lewis}, \&
  {Okamoto}}]{bailin-etal05-diskhalo}
{Bailin}, J., {Kawata}, D., {Gibson}, B.~K., et~al. 2005, \apj, 627, L17

\bibitem[{{Bell} {et~al.}(2010){Bell}, {Xue}, {Rix}, {Ruhland}, \&
  {Hogg}}]{bell-etal10}
{Bell}, E.~F., {Xue}, X.~X., {Rix}, H., {Ruhland}, C., \& {Hogg}, D.~W. 2010,
  \aj, 140, 1850

\bibitem[{{Bell} {et~al.}(2008){Bell}, {Zucker}, {Belokurov}, {Sharma},
  {Johnston}, {Bullock}, {Hogg}, {Jahnke}, {de Jong}, {Beers}, {Evans},
  {Grebel}, {Ivezi{\'c}}, {Koposov}, {Rix}, {Schneider}, {Steinmetz}, \&
  {Zolotov}}]{bell-etal08}
{Bell}, E.~F., {Zucker}, D.~B., {Belokurov}, et~al. 2008, \apj, 680, 295

\bibitem[{{Bullock} \& {Johnston}(2005)}]{bj05}
{Bullock}, J.~S., \& {Johnston}, K.~V. 2005, \apj, 635, 931 (BJ05)

\bibitem[{{Cooper} {et~al.}(2010){Cooper}, {Cole}, {Frenk}, {White}, {Helly},
  {Benson}, {De Lucia}, {Helmi}, {Jenkins}, {Navarro}, {Springel}, \&
  {Wang}}]{cooper-etal10}
{Cooper}, A.~P., {Cole}, S., {Frenk}, C.~S., et~al. 2010, \mnras, 406, 744 (C10)

\bibitem[{{Debattista} {et~al.}(2008){Debattista}, {Moore}, {Quinn},
  {Kazantzidis}, {Maas}, {Mayer}, {Read}, \& {Stadel}}]{debattista08}
{Debattista}, V.~P., {Moore}, B., {Quinn}, T., et~al. 2008, \apj, 681, 1076

\bibitem[{{Font} {et~al.}(2006){Font}, {Johnston}, {Bullock}, \&
  {Robertson}}]{font-etal06}
{Font}, A.~S., {Johnston}, K.~V., {Bullock}, J.~S., \& {Robertson}, B.~E. 2006,
  \apj, 638, 585

\bibitem[{{Font} {et~al.}(2011){Font}, {McCarthy}, {Crain}, {Theuns}, {Schaye},
  {Wiersma}, \& {Dalla Vecchia}}]{font-etal11}
{Font}, A.~S., {McCarthy}, I.~G., {Crain}, R.~A., et~al. 2011, \mnras, 416, 2802

\bibitem[{{Gao} {et~al.}(2011){Gao}, {Frenk}, {Boylan-Kolchin}, {Jenkins},
  {Springel}, \& {White}}]{gao11}
{Gao}, L., {Frenk}, C.~S., {Boylan-Kolchin}, M., et~al. 2011, \mnras, 410, 2309

\bibitem[{{Helmi} {et~al.}(2011){Helmi}, {Cooper}, {White}, {Cole}, {Frenk}, \&
  {Navarro}}]{helmi-etal11}
{Helmi}, A., {Cooper}, A.~P., {White}, S.~D.~M., et~al. 2011, \apjl, 733, L7

\bibitem[{{Ibata} {et~al.}(2013){Ibata}, {Lewis}, {McConnachie}, {Martin},
  {Irwin}, {Ferguson}, {Babul}, {Bernard}, {Chapman}, {Collins}, {Fardal},
  {Mackey}, {Navarro}, {Penarrubia}, {Rich}, {Tanvir}, \& {Widro}}]{ibata13}
{Ibata}, R.~A., {Lewis}, G.~F., {McConnachie}, A.~W., et~al. 2013, ApJ, in press, arXiv:1311.5888

\bibitem[{{Kazantzidis} {et~al.}(2008){Kazantzidis}, {Bullock}, {Zentner},
  {Kravtsov}, \& {Moustakas}}]{kazantzidis08}
{Kazantzidis}, S., {Bullock}, J.~S., {Zentner}, A.~R., {Kravtsov}, A.~V., \&
  {Moustakas}, L.~A. 2008, \apj, 688, 254

\bibitem[{{Kazantzidis} {et~al.}(2004){Kazantzidis}, {Kravtsov}, {Zentner},
  {Allgood}, {Nagai}, \& {Moore}}]{kazantzidis-etal04}
{Kazantzidis}, S., {Kravtsov}, A.~V., {Zentner}, A.~R., et~al. 2004, \apj, 611, L73

\bibitem[{{Knebe} \& {Wie{\ss}ner}(2006)}]{knebe06}
{Knebe}, A., \& {Wie{\ss}ner}, V. 2006, \pasa, 23, 125

\bibitem[{{Knollmann} \& {Knebe}(2009)}]{AHF}
{Knollmann}, S.~R., \& {Knebe}, A. 2009, \apjs, 182, 608

\bibitem[{{Koposov} {et~al.}(2009){Koposov}, {Yoo}, {Rix}, {Weinberg},
  {Macci{\`o}}, \& {Escud{\'e}}}]{koposov09}
{Koposov}, S.~E., {Yoo}, J., {Rix}, H.-W., et~al. 2009, \apj, 696, 2179

\bibitem[{{Kravtsov}(2010)}]{kravtsov10}
{Kravtsov}, A. 2010, {Advances in Astronomy}, 2010, 281913,
  doi:10.1155/2010/281913

\bibitem[{{Libeskind} {et~al.}(2011){Libeskind}, {Knebe}, {Hoffman},
  {Gottl{\"o}ber}, \& {Yepes}}]{libeskind-etal11}
{Libeskind}, N.~I., {Knebe}, A., {Hoffman}, Y., {Gottl{\"o}ber}, S., \&
  {Yepes}, G. 2011, \mnras, 418, 336 (L11)

\bibitem[{{Majewski} {et~al.}(2003){Majewski}, {Skrutskie}, {Weinberg}, \&
  {Ostheimer}}]{majewski-etal03}
{Majewski}, S.~R., {Skrutskie}, M.~F., {Weinberg}, M.~D., \& {Ostheimer}, J.~C.
  2003, \apj, 599, 1082

\bibitem[{{McConnachie} {et~al.}(2009){McConnachie}, {Irwin}, {Ibata},
  {Dubinski}, {Widrow}, {Martin}, {C{\^o}t{\'e}}, {Dotter}, {Navarro},
  {Ferguson}, {Puzia}, {Lewis}, {Babul}, {Barmby}, {Bienaym{\'e}}, {Chapman},
  {Cockcroft}, {Collins}, {Fardal}, {Harris}, {Huxor}, {Mackey},
  {Pe{\~n}arrubia}, {Rich}, {Richer}, {Siebert}, {Tanvir}, {Valls-Gabaud}, \&
  {Venn}}]{mcconnachie-etal09}
{McConnachie}, A.~W., {Irwin}, M.~J., {Ibata}, R.~A., et~al. 2009, \nat, 461, 66

\bibitem[{{Monachesi} {et~al.}(2013){Monachesi}, {Bell}, {Radburn-Smith},
  {Vlaji{\'c}}, {de Jong}, {Bailin}, {Dalcanton}, {Holwerda}, \&
  {Streich}}]{monachesi13}
{Monachesi}, A., {Bell}, E.~F., {Radburn-Smith}, D.~J., et~al. 2013, \apj, 766, 106

\bibitem[{{Nickerson} {et~al.}(2011){Nickerson}, {Stinson}, {Couchman},
  {Bailin}, \& {Wadsley}}]{nickerson11}
{Nickerson}, S., {Stinson}, G., {Couchman}, H.~M.~P., {Bailin}, J., \&
  {Wadsley}, J. 2011, \mnras, 415, 257

\bibitem[{{Pe{\~n}arrubia} {et~al.}(2010){Pe{\~n}arrubia}, {Benson}, {Walker},
  {Gilmore}, {McConnachie}, \& {Mayer}}]{penarrubia10_cusp}
{Pe{\~n}arrubia}, J., {Benson}, A.~J., {Walker}, M.~G., et~al. 2010, \mnras, 406, 1290

\bibitem[{{Purcell} {et~al.}(2007){Purcell}, {Bullock}, \&
  {Zentner}}]{purcell-etal07}
{Purcell}, C.~W., {Bullock}, J.~S., \& {Zentner}, A.~R. 2007, \apj, 666, 20

\bibitem[{{Rashkov} {et~al.}(2012){Rashkov}, {Madau}, {Kuhlen}, \&
  {Diemand}}]{rashkov12}
{Rashkov}, V., {Madau}, P., {Kuhlen}, M., \& {Diemand}, J. 2012, \apj, 745, 142

\bibitem[{{Schlaufman} {et~al.}(2012){Schlaufman}, {Rockosi}, {Lee}, {Beers},
  {Allende Prieto}, {Rashkov}, {Madau}, \& {Bizyaev}}]{schlaufman12}
{Schlaufman}, K.~C., {Rockosi}, C.~M., {Lee}, Y.~S., et~al. 2012, \apj, 749, 77

\bibitem[{{Spergel} {et~al.}(2007){Spergel}, {Bean}, {Dor{\'e}}, {Nolta},
  {Bennett}, {Dunkley}, {Hinshaw}, {Jarosik}, {Komatsu}, {Page}, {Peiris},
  {Verde}, {Halpern}, {Hill}, {Kogut}, {Limon}, {Meyer}, {Odegard}, {Tucker},
  {Weiland}, {Wollack}, \& {Wright}}]{WMAP3}
{Spergel}, D.~N., {Bean}, R., {Dor{\'e}}, O., et~al. 2007, \apjs, 170, 377

\bibitem[{{Stadel} {et~al.}(2009){Stadel}, {Potter}, {Moore}, {Diemand},
  {Madau}, {Zemp}, {Kuhlen}, \& {Quilis}}]{stadel09}
{Stadel}, J., {Potter}, D., {Moore}, B., et~al. 2009, \mnras, 398, L21

\bibitem[{{Stinson} {et~al.}(2010){Stinson}, {Bailin}, {Couchman}, {Wadsley},
  {Shen}, {Nickerson}, {Brook}, \& {Quinn}}]{MUGS}
{Stinson}, G.~S., {Bailin}, J., {Couchman}, H., et~al. 2010, \mnras, 408, 812

\bibitem[{{Tasca} \& {White}(2011)}]{tasca11}
{Tasca}, L.~A.~M., \& {White}, S.~D.~M. 2011, \aap, 530, A106

\bibitem[{{Tissera} {et~al.}(2013){Tissera}, {Scannapieco}, {Beers}, \&
  {Carollo}}]{tissera13}
{Tissera}, P.~B., {Scannapieco}, C., {Beers}, T.~C., \& {Carollo}, D. 2013,
  \mnras, 432, 3391

\bibitem[{{Tissera} {et~al.}(2010){Tissera}, {White}, {Pedrosa}, \&
  {Scannapieco}}]{tissera10}
{Tissera}, P.~B., {White}, S.~D.~M., {Pedrosa}, S., \& {Scannapieco}, C. 2010,
  \mnras, 406, 922

\bibitem[{{Tremonti} {et~al.}(2004){Tremonti}, {Heckman}, {Kauffmann},
  {Brinchmann}, {Charlot}, {White}, {Seibert}, {Peng}, {Schlegel}, {Uomoto},
  {Fukugita}, \& {Brinkmann}}]{tremonti-etal04}
{Tremonti}, C.~A., {Heckman}, T.~M., {Kauffmann}, G., et~al. 2004, \apj, 613, 898

\bibitem[{{Valluri} {et~al.}(2010){Valluri}, {Debattista}, {Quinn}, \&
  {Moore}}]{valluri10}
{Valluri}, M., {Debattista}, V.~P., {Quinn}, T., \& {Moore}, B. 2010, \mnras,
  403, 525

\bibitem[{{Xue} {et~al.}(2011){Xue}, {Rix}, {Yanny}, {Beers}, {Bell}, {Zhao},
  {Bullock}, {Johnston}, {Morrison}, {Rockosi}, {Koposov}, {Kang}, {Liu},
  {Luo}, {Lee}, \& {Weaver}}]{xue11}
{Xue}, X.-X., {Rix}, H.-W., {Yanny}, B., et~al. 2011, \apj, 738, 79

\bibitem[{{Zemp} {et~al.}(2011){Zemp}, {Gnedin}, {Gnedin}, \&
  {Kravtsov}}]{zemp11}
{Zemp}, M., {Gnedin}, O.~Y., {Gnedin}, N.~Y., \& {Kravtsov}, A.~V. 2011, \apjs,
  197, 30

\bibitem[{{Zolotov} {et~al.}(2009){Zolotov}, {Willman}, {Brooks}, {Governato},
  {Brook}, {Hogg}, {Quinn}, \& {Stinson}}]{zolotov09}
{Zolotov}, A., {Willman}, B., {Brooks}, A.~M., et~al. 2009, \apj, 702, 1058

\end{thebibliography}

\end{document}